\documentclass[twocolumn]{aastex701}
\usepackage{amsmath}
\usepackage{url}
\usepackage{mathrsfs}
\shorttitle{Shock Acceleration in GRB Afterglows}
\shortauthors{Wu et al.}

\begin{document}
\title{Maximum Energy of Particles Accelerated in Gamma-Ray Burst Afterglow Shocks}

\author[orcid=0000-0002-0196-9169]{Zhao-Feng Wu}
\affiliation{Department of Physics and Astronomy, Purdue University, 525 Northwestern Avenue, West Lafayette, IN 47907, USA}
\email[show]{wu2177@purdue.edu}

\author[orcid=0000-0002-2193-3809]{Sofía Guevara-Montoya}
\affiliation{Departamento de Física, Universidad Nacional de Colombia, Bogotá 111321, Colombia}
\email[]{soguevaram@unal.edu.co}

\author[orcid=0000-0001-7833-1043]{Paz Beniamini}
\affiliation{Department of Natural Sciences, The Open University of Israel, P.O. Box 808, Ra’anana 4353701, Israel}
\affiliation{Astrophysics Research Center of the Open University (ARCO), The Open University of Israel, P.O. Box 808, Ra’anana 4353701, Israel}
\affiliation{Department of Physics, The George Washington University, 725 21st Street NW, Washington, DC 20052, USA}
\email[]{paz.beniamini@gmail.com}

\author[orcid=0000-0003-1503-2446]{Dimitrios Giannios}
\affiliation{Department of Physics and Astronomy, Purdue University, 525 Northwestern Avenue, West Lafayette, IN 47907, USA}
\email[]{dgiannio@purdue.edu}

\author[orcid=0000-0002-5408-3046]{Daniel Grošelj}
\affiliation{Centre for mathematical Plasma Astrophysics, Department of Mathematics, KU Leuven, B-3001 Leuven, Belgium}
\email[]{daniel.groselj@kuleuven.be}

\author[orcid=0000-0002-1227-2754]{Lorenzo Sironi}
\affiliation{Department of Astronomy and Columbia Astrophysics Laboratory, Columbia University, New York, NY 10027, USA}
\affiliation{Center for Computational Astrophysics, Flatiron Institute, 162 5th Avenue, New York, NY 10010, USA}
\email[]{lsironi@astro.columbia.edu}

\begin{abstract}
\noindent Particle acceleration in relativistic collisionless shocks remains an open problem in high-energy astrophysics. Particle-in-cell (PIC) simulations predict that electron acceleration in weakly magnetized shocks proceeds via small-angle scattering, leading to a maximum electron energy significantly below the Bohm limit. This upper bound on electron energy manifests observationally as a characteristic synchrotron cutoff, providing a direct probe of the underlying acceleration physics. Gamma-ray burst (GRB) afterglows offer an exceptional laboratory for testing these predictions. Here, we model the spectral evolution of GRB afterglows during the relativistic deceleration phase, incorporating PIC-motivated acceleration prescriptions and self-consistently computing synchrotron and synchrotron self-Compton emission. We find that low-energy bursts in low-density environments, typical of short GRBs, exhibit a pronounced synchrotron cutoff in the GeV band within minutes to hours after the trigger. Applying our framework to GRB~190114C and GRB~130427A, we find that current observations are insufficient to discriminate between PIC-motivated acceleration and the Bohm limit, primarily due to poor photon statistics in the Fermi-LAT band. Nevertheless, future MeV–TeV afterglow observations can break model degeneracies and place substantially tighter constraints on the mechanisms responsible for particle acceleration in relativistic shocks. To this end, we simulate a fiducial nearby short GRB as a promising probe of the cutoff location, for which the two acceleration scenarios are cleanly distinguishable and the detection of such an event in the near future remains feasible.
\end{abstract}


\section{Introduction} \label{intro}
Relativistic collisionless shocks are among the most efficient particle accelerators in the Universe. Their signatures are observed across a wide range of astrophysical environments, yet the physical mechanisms governing particle acceleration in these systems remain poorly understood~\citep{2011MNRAS.412..522B,2012ApJ...749...80S,2015SSRv..191..519S}. One of the leading candidates is the diffusive shock acceleration process, in which particles gain energy through repeated crossings of the shock front mediated by magnetic turbulence \citep[e.g.,][]{1949PhRv...75.1169F,2011ApJ...726...75S,1978MNRAS.182..147B,1978ApJ...221L..29B,2001MNRAS.328..393A}. In the most optimistic scenario, the scattering mean free path approaches the Larmor radius of the accelerated particle, corresponding to the so-called Bohm limit~\citep{dejager92}, which represents the theoretical upper bound on acceleration rate.

The non-linear processes taking place at the shock front can be modeled from first principles using particle-in-cell (PIC) simulations, which have substantially advanced our understanding of particle acceleration at relativistic shocks~\citep[e.g.,][]{2008ApJ...682L...5S,2008ApJ...673L..39S,2009ApJ...695L.189M,2015SSRv..191..519S}. In weakly magnetized environments, simulations show that particle acceleration proceeds via small-angle scattering, leading to a slower increase of the maximum electron Lorentz factor, $\gamma_{\max} \propto t^{1/2}$, rather than the linear growth expected in the Bohm limit \citep{2009ApJ...693L.127K,2010ApJ...710L..16K,2013MNRAS.430.1280P,2013ApJ...771...54S,2024ApJ...963L..44G}. 

Gamma-ray bursts (GRBs) provide an exceptional laboratory for testing particle acceleration predictions. In the standard afterglow framework, the interaction between the relativistic ejecta and the circumburst medium results into a forward shock that accelerates electrons, producing broadband emission via synchrotron and synchrotron self-Compton (SSC) radiation \citep[e.g.,][]{1998ApJ...497L..17S,2001ApJ...548..787S,2022Galax..10...66M}. When radiative losses are taken into account, particle acceleration is limited to a maximum Lorentz factor, leading to a characteristic cutoff in the synchrotron spectrum \citep{2013ApJ...771...54S}. For typical GRB parameters, this cutoff is expected to appear in the GeV band during the early afterglow phase. Detecting such a feature would therefore provide a direct and robust probe of the microphysics governing particle acceleration in relativistic shocks.

Observationally, however, identifying the synchrotron cutoff remains challenging. Although Fermi-LAT has enabled systematic detections of long-lived GeV emission from GRB afterglows, a clear synchrotron cutoff has not been observed yet~\citep{2019ApJ...878...52A}, likely due to dominance of a rising SSC component. In such a case, the observed spectrum does not necessarily exhibit a sharp cutoff, but instead shows a spectral hardening around $\mathrm{GeV}$~\citep{2017ApJ...837...13P,2022A&A...657A.111H}. Instrumental sensitivity and photon statistics limitations in the GeV band further hinder the identification of both cutoff and spectral hardening features~\citep{2019ApJ...878...52A}.

Recent detections of very-high-energy (VHE; $\gtrsim 300$~GeV) emission from GRB afterglows by ground-based Cherenkov telescopes, such as MAGIC~\citep{2016APh....72...61A,2016APh....72...76A}, directly sample the SSC component and help break degeneracies in spectral modeling~\citep{MAGIC:2019irs}, enabling more robust observational probes of the particle acceleration timescales. By combining GeV and TeV observations with X-ray data, one can place significantly tighter constraints on the maximum synchrotron photon energy and, consequently, on the underlying particle acceleration physics.

In this paper, we test PIC-motivated particle acceleration models for weakly magnetized collisionless shocks in the context of GRB afterglow observations. We implement acceleration prescriptions motivated by the PIC simulations and by the Bohm diffusion limit within a self-consistent numerical framework for modeling GRB afterglow spectra. We show that low-energy bursts in low-density environments are most likely to yield a detectable synchrotron cutoff. Applying our model to GRB~190114C and GRB~130427A, we assess current observational constraints and discuss prospects for future detections. We also present a simulated nearby, on-axis short GRB as a promising probe of the cutoff location, and provide an estimate of its detection rate.

\section{Methodology} \label{Sec2}
We first discuss particle acceleration models and their associated spectral cutoffs for relativistic shocks in general, and then incorporate them in the context of GRB afterglows. Throughout this work, primed quantities denote values measured in the post-shock fluid comoving frame, lowercase unprimed quantities refer to the central engine frame, and uppercase quantities correspond to the observer frame, unless stated otherwise. Gaussian-cgs units are used throughout.

\subsection{Particle Acceleration Prescription} \label{Sec_acc}
Weakly magnetized relativistic shocks are expected to accelerate electrons into a nonthermal power-law distribution (e.g., \citealt{2001MNRAS.328..393A}) of the form
\begin{equation}
    n'(\gamma) \propto \gamma^{-p}
\end{equation}
valid over the range $\gamma_{\min} \leq \gamma \leq \gamma_{\max}$, where $p$ is the power-law index and $\gamma$ is the electron Lorentz factor measured in the frame comoving with the shocked fluid. Hereafter, all electron Lorentz factors $\gamma$ are defined in the comoving frame. Assuming a pure power-law electron distribution, the minimum Lorentz factor is determined by the shock jump conditions \citep{1996ApJ...473..204S} and can be written as
\begin{equation} \label{gamma_min}
    \gamma_{\min } \simeq \varepsilon_{\mathrm{e}} \frac{p-2}{p-1} \frac{m_{\mathrm{p}}}{m_{\mathrm{e}}} \Gamma,
\end{equation}
where $\varepsilon_{\mathrm{e}}$ is the fraction of post-shock internal energy carried by electrons, $m_{\mathrm{e}}$ and $m_{\mathrm{p}}$ are the electron and proton masses, respectively, and $\Gamma$ is the bulk Lorentz factor of the shocked fluid. The expression is valid in the limit $\gamma_{\min}\ll\gamma_{\max}$ with typical spectral indices $p\simeq2.4$.\footnote{For $p \simeq 2$, the dependence of $\gamma_{\min}$ on the high-energy cutoff becomes non-negligible. In our calculations, we use the full expression that accounts for both $\gamma_{\max}$ and $p$, although the difference is negligible for the parameter values considered here.}

PIC simulations performed in the past couple of decades indicate that electrons accelerated at relativistic shocks form a nonthermal power-law distribution with index $p \simeq 2.4$ and an energy fraction $\varepsilon_{\mathrm{e}} \sim 0.1$ for weakly magnetized shocks~\citep[e.g.,][]{2013ApJ...771...54S}. In addition, studies of GRB afterglows suggest that $\varepsilon_{\mathrm{e}} \sim 0.1-0.15$ with little scatter \citep{2014MNRAS.443.3578N,2017MNRAS.472.3161B}. More recent PIC studies suggest the possible emergence of an additional structure in the particle spectrum, including a suprathermal component, toward the end of the simulation \citep{2024ApJ...963L..44G}. A detailed exploration of these effects is beyond the scope of this work. Throughout this study, we adopt $\varepsilon_e = 0.1$ and assume a power-law index of $p \simeq 2.4$.

The maximum electron Lorentz factor $\gamma_{\max}$ is determined by balancing the acceleration rate against synchrotron cooling losses, as SSC losses are strongly suppressed in the deep Klein–Nishina (KN) regime. We consider two representative particle acceleration prescriptions, which differ in their acceleration rates and therefore predict different maximum attainable energies. 

The first acceleration model is motivated by PIC simulations \cite[e.g.,][]{2013ApJ...771...54S,2018MNRAS.477.5238P,2024ApJ...963L..44G}, which indicate small-angle scattering at the shock front. In this framework, the maximum electron Lorentz factor grows with time as $\gamma_{\rm max, PIC}(t') \propto (\omega_{\mathrm{p}}t')^{1/2}$, and the acceleration rate can be written as
\begin{equation}
    \gamma_{\rm max, PIC}\frac{d \gamma_{\rm max, PIC}}{d t'} = 0.25\, \left( \frac{m_{\rm p}}{m_{\rm e}} \right)^2\omega_{\mathrm{p}}\, \Gamma^2 \varepsilon_B\left( \frac{\lambda' \omega_{\mathrm{p}}}{c}\right),
\end{equation}
where the $(m_{\rm p}/m_{\rm e})^2$ factor reflects the assumption, motivated by PIC simulations, that the maximum electron energy can keep up with the maximum ion energy if cooling is neglected. Here $\lambda'$ is the transverse coherence length of the magnetic field, $\omega_{\mathrm{p}}$ is the frame-independent proton plasma frequency of the ambient medium \citep{2007A&A...475....1A},
\begin{equation}
    \omega_{\mathrm{p}} \equiv \sqrt{4 \pi e^2 n/m_p} \simeq 1.3 \times 10^3 n_0^{1 / 2} \, \mathrm{Hz},
\end{equation}
where $n = n_0 \, {\rm cm}^{-3}$, and $\varepsilon_B$ is the magnetic energy fraction\footnote{We adopt the same definition of $\varepsilon_B$ as in \citet{2013ApJ...771...54S} and \citet{2024ApJ...963L..44G}.} defined as
\begin{equation}
    \varepsilon_{\text B} \equiv \frac{B'^2 }{8 \pi \Gamma^2 n m_{\mathrm{p}} c^2}.
\end{equation}
Recent simulations \citep{2024ApJ...963L..44G} suggest that $\varepsilon_B(\lambda' \omega_{\mathrm{p}}/c) \simeq 0.4$, although this combination may be increasing for longer simulations. When synchrotron cooling is included, the evolution of the maximum electron Lorentz factor is governed by
\begin{equation}
\frac{d \gamma_{\rm max, PIC}}{d t'} \simeq \left( \frac{m_{\rm p}}{m_{\rm e}} \right)^2 \frac{0.1 \omega_{\mathrm{p}} \Gamma^2}{\gamma_{\rm max, PIC}} - \frac{\sigma_\mathrm{T} \gamma_{\rm max, PIC}^2}{6\pi} \frac{B'^2}{m_{\rm e} c}.
\end{equation}
All quantities are evaluated in the comoving frame of the shocked fluid, with $t'$ denoting comoving time and $B'$ the comoving magnetic field strength. Balancing acceleration and synchrotron cooling yields an equilibrium Lorentz factor and represents the maximum attainable electron energy, 
\begin{equation} \label{gamma_eq}
    \gamma_\mathrm{max, PIC} = \left( \frac{0.6 \pi m_{\rm p}^2 \omega_{\mathrm{p}} \Gamma^2 c}{\sigma_\mathrm{T} m_{\rm e} B'^2} \right)^{1/3}  = 1.3 \times 10^7  n_0^{-1/6} \varepsilon_{B,-2.5}^{-1/3},
\end{equation}
where $\varepsilon_{\text B} = 10^{-2.5} \varepsilon_{\text B,-2.5}$. 

For relativistic outflows with $\Gamma \gtrsim 100$, $\gamma_\mathrm{max}$ is typically reached within seconds in the observer frame. An analytic estimate for the acceleration timescale is
\begin{equation}
    T_{\rm PIC} \simeq \frac{\gamma_\mathrm{max, PIC}}{\Gamma} \left( \frac{0.1 m_{\rm p}^2 \omega_{\mathrm{p}} \Gamma^2}{m_{\rm e}^2\gamma_\mathrm{max, PIC}} \right)^{-1} = 0.4 \,\Gamma_{2}^{-3} n_0^{-5/6}  \varepsilon_{B,-2.5}^{-2/3} \,\mathrm{s},
\end{equation}
where $\Gamma = 100\Gamma_{2}$. This upper cutoff in the electron distribution should manifest as an exponential cutoff in the observed synchrotron spectrum. We can compute the corresponding characteristic synchrotron frequency for $\gamma_\mathrm{max}$, which serves as the beginning of the cutoff~\citep{1998ApJ...497L..17S},
\begin{equation}
    h \nu_{\mathrm{max,PIC}} =  \frac{h \Gamma e B'\gamma_\mathrm{max,PIC}^2 }{2 \pi m_e c} \simeq 0.2 \,\Gamma_{2}^{2} \, n_{0}^{1/6} \varepsilon_{B,-2.5}^{-1 / 6}\, \mathrm{GeV}.
\end{equation}

The expressions for $\gamma_\mathrm{max,PIC}$ and $h\nu_{\rm max,PIC}$ derived above depend on the magnetic field properties in the vicinity of the shock, in particular the magnetic energy fraction $\varepsilon_B$ and the coherence scale $\lambda'$, which regulate particle diffusion and cooling losses. However, extrapolating these parameters from the simulation scales to the astrophysical scales remains uncertain. Moreover, some studies indicate that the magnetic field may decay downstream of the shock front~\citep{2008IJMPD..17.1769C,2012MNRAS.427L..40K,2013MNRAS.428..845L,2015JPlPh..81a4501L,2015MNRAS.453.3772L}, implying that the $\varepsilon_B$ averaged over the full emission region can be lower than its value immediately behind the shock. The highest-energy particles predominantly cool close to the shock, while lower-energy particles are more sensitive to the averaged $\varepsilon_B$ over the whole emission region.

While the evolution of magnetic field properties could in principle be treated with a multi-zone model, both $\gamma_\mathrm{max,PIC}$ and $h\nu_{\rm max,PIC}$ depend only weakly on $\varepsilon_B$. Therefore, the high-energy cutoff properties remain largely insensitive to moderate variations in $\varepsilon_B$, provided that the combination $\varepsilon_B(\lambda' \omega_{\mathrm{p}}/c) \simeq 0.4$ is preserved. This motivates the use of an effective one-zone description. Consequently, and to remain consistent with PIC simulation results, we allow $\varepsilon_B$ to vary only modestly towards lower values relative to our reference value of $\langle \varepsilon_B \rangle \simeq 3.5 \times 10^{-3}$. Specifically, we explore the range $10^{-4} \lesssim \varepsilon_B \lesssim 5 \times 10^{-3}$ while keeping $\varepsilon_B(\lambda' \omega_{\mathrm{p}}/c) \simeq 0.4$ fixed, following a prescription similar to that adopted by \citet{2013ApJ...771...54S}. Such values are also consistent with inferences from GRB afterglow modeling \citep{2014ApJ...785...29S,2015MNRAS.454.1073B}. A full treatment using a multi-zone model is left for future work.

For Bohm diffusion \citep{dejager92}, electrons gain energy by a factor of order unity per gyration, corresponding to large-angle scattering that is significantly faster than small-angle scattering predicted by PIC simulations. Following the same balance between acceleration and synchrotron cooling, the maximum Lorentz factor becomes
\begin{equation}
    \gamma_\mathrm{max, Bohm} = \sqrt{\frac{6 \pi e}{\sigma_T B^{\prime}}} = 1.1 \times 10^8 \varepsilon_{B,-2.5}^{-1/4} \Gamma_{2}^{-1/2} n_0^{-1/4}.
\end{equation}
In this case, $\gamma_{\mathrm{max, Bohm}}$ increases as the jet decelerates. The corresponding acceleration timescale and characteristic synchrotron frequency are
\begin{equation}
        T_{\rm Bohm} \simeq \frac{\gamma_\mathrm{max, Bohm}}{\Gamma} \left( \frac{m_{\rm e} c}{eB'} \right) = 5\, \Gamma_{2}^{-5/2} n_0^{-1/4}  \varepsilon_{B,-2.5}^{-3/4} \,\mathrm{s},
\end{equation}
and 
\begin{equation}
    h \nu_{\mathrm{max, Bohm}} =  \frac{h \Gamma e B'\gamma_{\rm max, Bohm}^2 }{2 \pi m_{\rm e} c} = 15.7 \,\Gamma_{2}\, \mathrm{GeV}.
\end{equation}
We note that $h \nu_{\mathrm{max, Bohm}}$ is independent of $n$ and $\varepsilon_{B}$ and remains fixed in the comoving frame of the shocked fluid. Bohm diffusion is considered to give a theoretical upper bound for $\gamma_{\max}$ and $h \nu_{\mathrm{max}}$, which is, however, not supported by present PIC simulations of relativistic weakly magnetized shocks. The ratio of the cutoff energies predicted by the two prescriptions is
\begin{equation}
    \frac{h \nu_{\mathrm{max, PIC}}}{h \nu_{\mathrm{max, Bohm}}} \simeq 0.013 \,\Gamma_{2} \, n_{0}^{1/6} \varepsilon_{B,-2.5}^{-1 / 6}
\end{equation}
indicating a difference of roughly two orders of magnitude that decreases for faster outflows, that depends linearly on the bulk Lorentz factor and very weakly on other physical parameters. 

\subsection{Dynamics of GRB Shocks} \label{Sec_hyd}
Powerful relativistic shocks in GRBs are generated as the ejecta from the central engine interact with the ambient medium. Here we introduce the dynamics of GRB shocks relevant for modeling GRB afterglow spectra. 

The relativistic ejecta initially propagate into the ambient medium with a constant Lorentz factor $\Gamma_0$. After sweeping up an external mass of order $M_0/\Gamma_0$, where $M_0$ is the initial ejecta mass, the shell begins to decelerate and approaches the self-similar solution described by~\citet{1976PhFl...19.1130B}. In this work, we assume that the shock has already entered the self-similar deceleration phase and neglect the contribution of the reverse shock. Although the reverse shock can affect the dynamics prior to the deceleration radius $r_{\mathrm{dec}}$~\citep{1995ApJ...455L.143S}, its influence is confined to the very early evolution and is therefore expected to have a limited impact on the considerations here.

We consider two external density profiles: a constant-density interstellar medium (ISM) and a wind environment. Under the assumption of adiabatic expansion into a constant-density ISM, the bulk Lorentz factor of the shocked fluid\footnote{This refers to the Lorentz factor of the shocked material immediately behind the shock front; the shock Lorentz factor itself is $\Gamma_{\rm sh}=\sqrt{2}\,\Gamma$.} during the relativistic deceleration phase follows the self-similar solution of \citet{1976PhFl...19.1130B},
\begin{equation}
    \Gamma(r)=\left(\frac{17 E_{\mathrm{iso}}}{16 \pi n m_p c^2}\right)^{\frac{1}{2}} r^{-\frac{3}{2}},
\end{equation}
valid at radii larger than the deceleration radius
\begin{equation}
    r_{\mathrm{dec}} \equiv \left( \frac{17 E_{\mathrm{iso}}}{16 \pi \Gamma_0^2 n m_p c^2}\right)^{\frac{1}{3}}
\end{equation}
where $E_{\mathrm{iso}}$ is the isotropic-equivalent explosion kinetic energy, $\Gamma_0$ is the initial Lorentz factor of the outflow, and $n$ is the ambient number density measured in the rest frame of the ISM. Writing $E_{\mathrm{iso}} = 10^{54}\, E_{{\mathrm{iso}},54} \,\mathrm{erg}$ and $\Gamma_0 = 10^{2.5}\, \Gamma_{0,2.5}$, the deceleration radius becomes
\begin{equation}
    r_{\mathrm{dec}} \simeq 1.3 \times 10^{17} \, \Gamma_{0,2.5}^{-2 / 3} \, E_{{\mathrm{iso}},54}^{1 / 3} \,  n_0^{-1 / 3} \mathrm{~cm} .
\end{equation}
We fix the initial Lorentz factor of the outflow at $\Gamma_0 = 500$ throughout, since our results are insensitive to this choice provided that the jet has entered the deceleration phase by the time of observation. During the relativistic deceleration phase, the observed emission is shaped by relativistic effects encapsulated in the Doppler factor,
\begin{equation}
    \mathcal{D} = \frac{1}{\Gamma(1 - \beta\cos\theta)},
\end{equation}
where $\beta$ is the dimensionless speed of the emitting region and $\theta$ is the angle between the emitting region and the observer’s line of sight. For on-axis emission, $\mathcal{D} \simeq 2\Gamma$, while emission from regions with $\theta \gtrsim 1/\Gamma$ is strongly suppressed by relativistic beaming. After averaging over the visible emitting region of angular size $\sim 1/\Gamma$, the effective Doppler factor is well approximated by $\mathcal{D} \simeq \Gamma$~\citep{2010ApJ...718L..63P}. The observed and comoving timescales are then related by
\begin{equation} \label{timescales}
    \mathrm{d} T=\frac{(1+z) \mathrm{d} t}{ \Gamma^2} = \frac{(1+z) \mathrm{d} t'}{ \Gamma}.
\end{equation}
where $T$, $t$ and $t'$ denote the observer, lab-frame and comoving times, respectively.

We now incorporate the shock dynamics into the spectral cutoffs discussed in Section~\ref{Sec_acc}. For the PIC-motivated acceleration prescription, the maximum Lorentz factor $\gamma_\mathrm{max,PIC}$ is independent of the bulk Lorentz factor $\Gamma$ and therefore remains constant during the deceleration phase. In contrast, the maximum Lorentz factor under the Bohm diffusion limit evolves as
\begin{equation}
    \gamma_\mathrm{max, Bohm} =  7.7 \times 10^7 \varepsilon_{B,-2.5}^{-1/4} \, E_{{\mathrm{iso}},54}^{-1 / 16} \, n_0^{-3/16} \, T_{\mathrm{obs}, 2}^{3 / 16},
\end{equation}
where $T_{\mathrm{obs},2}$ is the observer time in units of 100~s. The corresponding characteristic synchrotron cutoff frequency for the PIC model is
\begin{equation} \label{nu_max}
h \nu_{\mathrm{max,PIC}} \simeq 0.8 \, \varepsilon_{B,-2.5}^{-1 / 6} \, E_{\mathrm{iso},54}^{1 / 4} \, n_{0}^{-1 / 12}  \,T_{\mathrm{obs}, 2}^{-3 / 4} (1+z)^{-1 / 4} \, \mathrm{GeV},
\end{equation}
while for the Bohm limit it becomes
\begin{equation} \label{nu_max_Bohm}
h \nu_{\mathrm{max, Bohm}} \simeq 31.5 \, E_{\mathrm{iso},54}^{1 / 8} \, n_{0}^{-1 / 8} \,T_{\mathrm{obs}, 2}^{-3 / 8} (1+z)^{-5 / 8} \, \mathrm{GeV}.
\end{equation}
The ratio of the two cutoff frequencies is then
\begin{equation} 
\frac{h \nu_{\mathrm{max, PIC}}}{h \nu_{\mathrm{max, Bohm}}} \simeq 0.025 \, \varepsilon_{B,-2.5}^{-1 / 6} \, E_{\mathrm{iso},54}^{1 / 8} \, n_{0}^{1 / 24} \,T_{\mathrm{obs}, 2}^{-3 / 8} (1+z)^{3 / 8},
\end{equation}
which depends weakly on the burst parameters and shows a modest evolution with observer time.


We also consider an adiabatic GRB blast wave propagating into the stellar wind of its progenitor, as expected for massive stars such as Wolf–Rayet stars (see \citealt{2007ARA&A..45..177C} for a review). For a standard constant-velocity wind, the density profile is
\begin{equation}
    \rho(r)=\frac{A}{r^2},
\end{equation}
where
\begin{equation}
    A=\frac{\dot{M}}{4 \pi v_{\mathrm{wind}}}.
\end{equation}
Here, $\dot{M}$ is the progenitor mass-loss rate and $v_{\text {wind }}$ is the wind velocity. For a canonical Wolf–Rayet star with $\dot{M}=10^{-5}\,M_\odot\,{\rm yr}^{-1}$ and $v_w=10^8\,{\rm cm\,s}^{-1}$~(see e.g., \citealt{2007ARA&A..45..177C}), one obtains $A \sim 5 \times 10^{11} \mathrm{~g} \mathrm{~cm}^{-1}$. We therefore write $A = 5 \times 10^{11} \, A_* \mathrm{~g} \mathrm{~cm}^{-1}$.\footnote{Afterglow modeling often infers wind densities below the canonical Wolf–Rayet value, in some cases even $A_* < 0.1$, suggesting that $A_* =1$ is toward the high end of the range relevant for GRB environments~\citep{2011A&A...526A..23S,2014ApJ...782....5H}.} The corresponding number density profile is parameterized as
\begin{equation}
n^{\rm wind}(r)=3\times10^{35}A_* r^{-2}\ {\rm cm}^{-3}.
\end{equation}
In this case, the bulk Lorentz factor of the shocked fluid evolves as
\begin{equation}
    \Gamma(r)=\left(\frac{9 E_{\mathrm{iso}}}{16 \pi A c^2}\right)^{1 / 2} r^{-1 / 2},
\end{equation}
starting from the deceleration radius 
\begin{equation}
    r_{\mathrm{dec}} \simeq 4 \times 10^{15} E_{\mathrm{iso},54} \Gamma_{0,2.5}^{-2} A_{*}^{-1} \mathrm{~cm}.
\end{equation}

For the PIC-motivated particle acceleration model, the maximum electron Lorentz factor and the corresponding synchrotron cutoff frequency are, respectively,
\begin{equation}
    \gamma^{\rm wind}_\mathrm{max, PIC} = 5.2 \times 10^6  \varepsilon_{B, -2.5}^{-1/3}  \, E_{{\mathrm{iso}},54}^{1/6} \, A_{*}^{-1/3} \, T_{\mathrm{obs}, 2}^{~1/6},
\end{equation}
and
\begin{equation}
    h \nu^{\rm wind}_{\mathrm{max,PIC}} \simeq 0.5 \,\varepsilon_{B, -2.5}^{-1/6}  \, E_{{\mathrm{iso}},54}^{1/3} \, A_{*}^{-1/6} \, T_{\mathrm{obs}, 2}^{-2/3}\, (1+z)^{-1 / 3} \,\mathrm{GeV}.
\end{equation}
For the Bohm limit, we obtain
\begin{equation}
    \gamma^{\rm wind}_\mathrm{max, Bohm} =  2.8 \times 10^7 \varepsilon_{B,-2.5}^{-1/4} \, E_{{\mathrm{iso}},54}^{~1 / 8} \, A_{*}^{-3/8} \, T_{\mathrm{obs}, 2}^{3 / 8},
\end{equation}
and
\begin{equation} 
h \nu^{\rm wind}_{\mathrm{max, Bohm}} \simeq 15 \, E_{{\mathrm{iso}},54}^{~1 / 4} \, A_{*}^{-1/4} \, T_{\mathrm{obs}, 2}^{-1 / 4} \, (1+z)^{-3 / 4} \,\mathrm{GeV}.
\end{equation}
The ratio of the cutoff frequencies is
\begin{equation}
    \frac{h \nu^{\rm wind}_{\mathrm{max, PIC}}}{h \nu^{\rm wind}_{\mathrm{max, Bohm}}} \simeq 0.03 \, \varepsilon_{B,-2.5}^{-1 / 6} \, E_{\mathrm{iso},54}^{1 / 12} \, A_{*}^{1/12} \,T_{\mathrm{obs}, 2}^{-5 / 12} (1+z)^{5 / 12},
\end{equation}
showing only a weak dependence on the burst parameters and a mild time evolution, similar to the ISM case. However, the particle density in the wind is significantly higher than in a typical ISM at observer times of minutes to hours,
\begin{equation}
    n^{\rm wind} \simeq 250 \, E_{\mathrm{iso},54}^{-1} \, A_{*}^{~2} T_{\mathrm{obs}, 2}^{-1} (1+z)\, \mathrm{cm}^{-3},
\end{equation}
which leads to a stronger SSC component that already dominates at $h \nu^{\rm wind}_{\mathrm{max}}$ (see Section~\ref{parameter}) and obscures the synchrotron cutoff. Hereafter, we focus on jets propagating into a low-density ISM-like medium rather than a dense wind environment. Although long GRBs are theoretically expected to reside in wind-like environments, afterglow modeling often favors constant-density media or density profiles shallower than the standard $\rho \propto r^{-2}$ wind~\citep{2006A&A...460..105V,2011A&A...526A..23S,2022MNRAS.511.2848A}, providing observational support for the ISM-like environments considered here.

The radial extent of the radiative zone requires more careful treatment. High-energy particles and pressure are concentrated within a thin shell immediately behind the shock front, whereas material advected farther downstream undergoes significant adiabatic expansion and experiences a decayed magnetic field, contributing negligibly to the high-energy emission. We therefore define the emitting region up to the similarity variable $\chi = 2$ in the Blandford–McKee solution~\citep{1976PhFl...19.1130B}. The corresponding shell thickness in the comoving frame of the shocked fluid is 
\begin{equation}
    \Delta^{\prime}(r)=\frac{r}{16 \Gamma},
\end{equation}
appropriate for a constant density medium.

Here, we adopt a single-zone approximation and assume that the bulk Lorentz factor of the emitting plasma is equal to that immediately behind the shock. In the comoving frame of the fluid just behind the shock, material at the downstream border of the emitting shell ($\chi = 2$) recedes at a speed of $c/3$. The particle escape timescale from the radiative region can therefore be expressed as
\begin{equation} \label{t_esc}
    t^{\prime}_{\text esc} = \frac{\Delta^{\prime}(r)}{c/3} = \frac{3r}{16 \Gamma c},
\end{equation}
and the comoving volume of the emitting region is then given by
\begin{equation} \label{V'}
    V^{\prime} = \pi \left(\frac{r}{\Gamma}\right)^2 \Delta^{\prime}(r) = \frac{\pi r^3}{16 \Gamma^3}.
\end{equation}

Here we do not account for the finite opening angle of the jet and instead approximate the outflow as spherical. This approximation is justified because the luminosity per unit solid angle is effectively identical to that of a spherical explosion as long as the relativistic beaming angle is smaller than the jet opening angle. The former increases over time, and the two eventually become equal at the `jet break time'~\citep[e.g.,][]{1997ApJ...487L...1R,1999ApJ...519L..17S}. Such a break occurs long after the early sub-GeV emission observed by Fermi-LAT, which is the regime of interest for identifying the maximum synchrotron photon energies.

\subsection{Kinetic Evolution and Radiation Processes} \label{Sec_kin}
We assume the shock properties are governed solely by the instantaneous Lorentz factor of the flow and model electron acceleration as instantaneous. The accelerated electrons are then advected downstream to radiate. Once electrons are injected into the radiative zone, the evolution is governed by the continuity equation
\begin{equation}
\frac{\partial N'(\gamma, t')}{\partial t'}=\frac{\partial}{\partial \gamma}[\dot{\gamma'}_{\rm cool} N'(\gamma, t')] + Q'_{\rm inj}(\gamma, t')-\frac{N'(\gamma, t')}{t'_{\mathrm{esc}}},
\end{equation}
where $N'(\gamma,t') = V'n'(\gamma,t')$ is the volume-integrated particle distribution in the comoving frame with the comoving volume $V'$ given by Equation~\eqref{V'}, and the escape timescale $t'_{\rm esc}$ is given by Equation~\eqref{t_esc}. The total cooling rate is
\begin{equation}
    \dot{\gamma}_{\rm {cool }}^{\prime}=\dot{\gamma}_{\rm {syn }}^{\prime}+\dot{\gamma}_{\rm {KN }}^{\prime} + \dot{\gamma}^{\prime}_{\mathrm{ad}},
\end{equation}
including synchrotron losses, SSC losses with KN corrections, and adiabatic cooling, respectively. We adopt the adiabatic loss term from \citet{2014A&A...564A..77P},
\begin{equation}
\dot{\gamma}_{\mathrm{ad}}^{\prime} = -\frac{3\gamma}{5t^{\prime}},
\end{equation}
which is appropriate for a constant-density ISM during the deceleration phase. The injection term $Q'_{\rm inj}$ represents the volume-integrated particle distribution of freshly accelerated electrons. The injected electron spectrum is assumed to follow a power law between $\gamma_{\min}$ and $\gamma_{\max}$, as defined in Section~\ref{Sec_acc}, with $\varepsilon_{\mathrm{e}}$ entering the spectral modeling only through $\gamma_{\min}$. Its normalization is set by the number of particles swept by the shock, given by the comoving particle flux $\Gamma n c$ through an area $\pi (r/\Gamma)^2$, such that
\begin{equation}
    \int_1^{\infty} Q'(\gamma, t') d \gamma = \Gamma n c\pi \left(\frac{r}{\Gamma}\right)^2 = \Gamma  n \pi  t'^2 c^3.
\end{equation} 

By formulating the problem in terms of the volume-integrated continuity equation, the effects of adiabatic dilution are naturally incorporated. An equivalent description can be obtained by introducing an effective escape term with a timescale comparable to the dynamical expansion time~\citep{2014A&A...564A..77P,2026MNRAS.546ag101A}, which is similar to our choice of $t'_{\rm esc}$.

The numerical calculations are performed using the code \textsc{Tleco}~\citep{2024ApJ...976..182D}, a one-zone model\footnote{The model assumes an isotropic distribution of particles with no spatial dependence.} that self-consistently evolves the particle distribution $N(\gamma,t')$ and computes the resulting synchrotron and SSC emission, including KN correction. To evaluate the synchrotron photon energy density relevant for SSC emission, we adopt a photon escape length equal to $\Delta'$, reflecting the thin-shell geometry of the emitting region. All physical ingredients required to evolve the particle spectrum and compute the radiative output have now been specified. We neglect internal $\gamma$–$\gamma$ absorption, as it primarily affects the VHE tail of the SSC component and is not central to our analysis.

The approximation of the Blandford–McKee solution by a one-zone model is not unique, and different choices for the effective emission region and characteristic timescales may introduce order of unity variations. However, such differences do not affect the qualitative behavior or the main conclusions of this work. We have verified that, when adopting identical assumptions and configurations, our numerical implementation reproduces the results of \citet{2014A&A...564A..77P}.

\section{Favored Burst Parameters} \label{parameter}
\begin{figure}
    \centering
    \includegraphics[width=\linewidth]{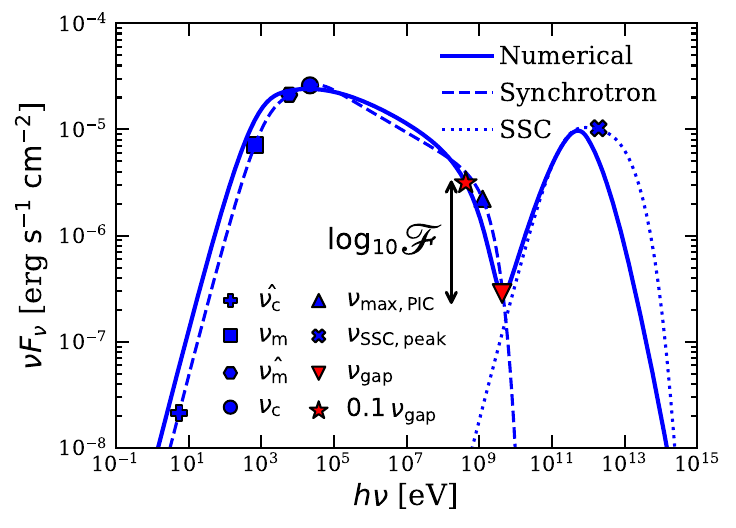}
    \vspace{-0.7cm}
    \caption{Afterglow spectrum for a burst placed at $D_{\mathrm{L}} = 100$~Mpc at $T_{\mathrm{obs}} = 100$~s. The burst parameters are $\varepsilon_e = 0.1$, $\varepsilon_B = 3.5 \times 10^{-3}$, $E_{\mathrm{iso}} = 10^{54}$~erg, $n = 0.5~\mathrm{cm^{-3}}$, and $p = 2.4$. The solid curve shows the numerical spectrum, while the dashed and dotted curves show the analytical synchrotron and SSC components with characteristic frequencies denoted by blue symbols. The red markers indicate $\nu_{\rm gap}$ and $0.1\,\nu_{\rm gap}$, with their vertical separation corresponding to $\log_{10}\mathscr{F}$, which quantifies the prominence of the synchrotron cutoff.}
    \label{fig:illustration}
    \vspace{-0.3cm}
\end{figure}

\begin{figure*} 
    \centering
    \vspace{-0.1cm} 
    \includegraphics[width=\linewidth]{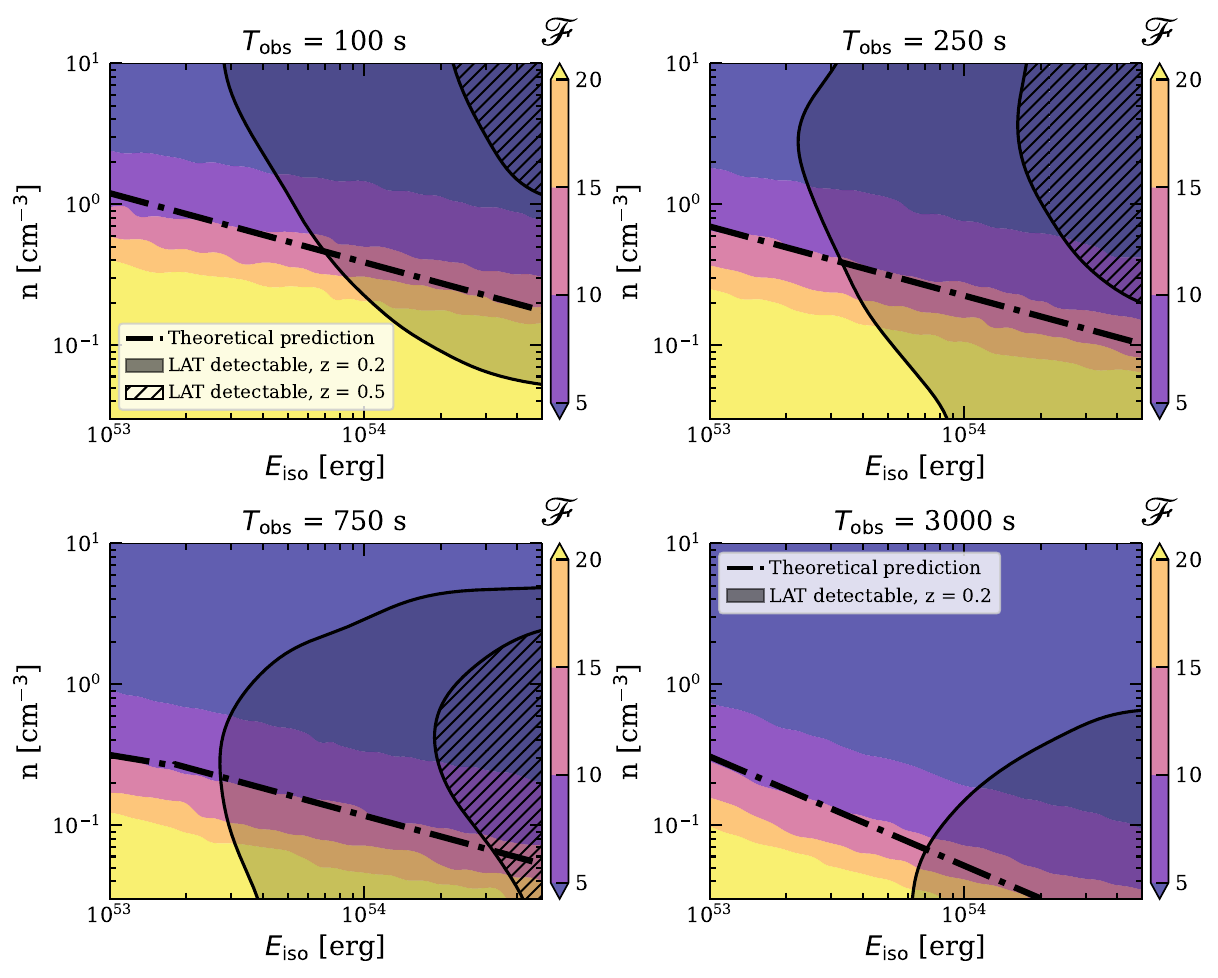}
    \vspace{-0.7cm} 
    \caption{Flux ratio $\mathscr{F} \equiv \nu F_\nu(0.1\,\nu_{\rm gap}) / \nu F_\nu(\nu_{\rm gap})$ shown as a function of $E_{\rm iso}$ and $n$ at observer times $T_{\rm obs} = 100,\ 250,\ 750\,$ and $3000\,\mathrm{s}$ after trigger. Colors indicate the value of $\mathscr{F}$, as shown by the color bar. All results are computed using the PIC-motivated acceleration prescription with $\varepsilon_e = 0.1$, $\varepsilon_B = 3.5 \times 10^{-3}$, and $p = 2.4$. Dash–dotted curves denote the theoretical prediction corresponding to $\mathscr{F} = 10$, in good agreement with the numerical results. The shaded and dashed contours indicate detectability for bursts placed at redshifts $z = 0.2$ and $z = 0.5$, respectively. }
    \label{E_n_plot}
\end{figure*}

The maximum electron energy manifests as an exponential cutoff in the GRB afterglow spectrum, as given by Equation~\eqref{nu_max}. Although this feature can be intrinsically sharp, it is often dominated by the rising SSC component, resulting in an apparent smooth spectral hardening. In such cases, the observed turnover provides only a lower bound on the maximum synchrotron frequency, yielding a weak constraint on particle acceleration. A significant detection of the exponential cutoff is possible only when it is not obscured by the SSC emission. In this section, we investigate the regions of GRB parameter space that favor a clear cutoff. All results presented in this section assume the PIC-motivated acceleration prescription for the synchrotron cutoff. For comparison, under the Bohm prescription the cutoff typically appears at much higher energies, where the emission is already dominated by the SSC component and is therefore rarely observable.

To quantify the cutoff feature, we define a flux ratio $\mathscr{F}$ designed to identify an exponential synchrotron drop, as opposed to a simple power-law decline, in observed spectra. We define
\begin{equation} \label{F_def}
\mathscr{F} = \frac{\nu F_\nu (0.1 \nu_{\rm gap})}{\nu F_\nu (\nu_{\rm gap})},
\end{equation}
where $\nu_{\rm gap}$ is the frequency at which $\nu F_\nu$ reaches its local minimum between the synchrotron and SSC peaks, marking the transition from synchrotron-dominated to SSC-dominated emission. 

As an illustration, Figure~\ref{fig:illustration} shows an example afterglow spectrum from our numerical calculations. For comparison, we also plot the analytical synchrotron and SSC components following~\citet{2024A&A...690A.281P}, along with the relevant characteristic frequencies, which are provided in Appendix~\ref{apped_1}. The numerical results are in good agreement with the analytical expectations. In the figure, the red $\triangledown$ marks the location of $\nu_{\rm gap}$ and the red star denotes $0.1\,\nu_{\rm gap}$, with their vertical separation on the logarithmic scale corresponding to $\log_{10}\mathscr{F}$. The ratio $\mathscr{F}$ therefore measures the effective spectral steepening immediately below the SSC-dominated regime, and becomes large when the synchrotron cutoff occurs well before the SSC component rises. Throughout, we adopt $\mathscr{F}>10$ as a practical criterion for a detectable exponential drop. For a pure synchrotron spectrum, the steepest physically relevant power-law decline is $\nu F_\nu \propto \nu^{(2-p)/2}$ \citep{1998ApJ...497L..17S}, which would require $p > 4$ to reproduce $\mathscr{F} > 10$ and is far steeper than expected from both theoretical considerations and observational constraints~\citep{2010ApJ...716L.135C}.

Figure~\ref{E_n_plot} shows $\mathscr{F}$ as a function of $E_{\rm iso}$ and $n$ at several observer times. Motivated by PIC simulations, we fix $\varepsilon_e=0.1$, $\varepsilon_B=3.5\times10^{-3}$, and $p=2.4$, and explore a parameter range representative of GRB afterglows, $10^{53}\,{\rm erg}<E_{\rm iso}<5\times10^{54}\,{\rm erg}$ and $0.03\,{\rm cm^{-3}}< n <10\,{\rm cm^{-3}}$~\citep{2018ApJ...859..160W,2022MNRAS.511.2848A,2023ApJ...959...13R}. We find that a large $\mathscr{F}$, corresponding to a more prominent synchrotron cutoff, is favored for low-energy explosions in low-density environments at early times. To further clarify the parameter dependence, we derive an analytic estimate
\begin{equation} \label{theoretical}
    \mathscr{F}  \simeq 2.2 \,\varepsilon_{\mathrm{e},-1}^{~1.3} \,\varepsilon_{B, -2.5}^{-0.1} \,E_{{\mathrm{iso}},54}^{-0.78} \,n_{0}^{-1.6} \,T_{\mathrm{obs}, 2}^{-0.95},
\end{equation}
evaluated for $p\simeq2.4$ at $T_{\mathrm{obs}}\simeq100$~s with $E_{\mathrm{iso}} \sim 10^{54}$~erg and $n \sim 1\,\mathrm{cm^{-3}}$. Expressions applicable to other parameter regimes and observer times, along with a detailed derivation, are given in Appendix~\ref{apped_2}. The dash–dotted curves in Figure~\ref{E_n_plot} show the corresponding theoretical prediction for $\mathscr{F}=10$ at different observer times, in good agreement with the numerical results.

Although low-energy bursts in low-density environments show a clear exponential cutoff, they may be intrinsically faint and therefore difficult to detect, depending on their distance. We therefore assess the detectability across the parameter space. Since $h\nu_{\mathrm{max}} \sim \mathrm{GeV}$, Fermi-LAT is well suited to probe this regime. A detailed prediction of detectability for a given spectral shape is highly instrument- and case-dependent and thus beyond the scope of this work. Instead, we estimate detectability using the observed cumulative fluence at $0.1\,\nu_{\rm gap}$ and evaluate the corresponding signal-to-noise ratio (SNR)\footnote{We adopt the Fermi-LAT instrument response functions from \url{https://www.slac.stanford.edu/exp/glast/groups/canda/lat_Performance.htm} with the mode of \texttt{P8R3\_SOURCE\_V3}. Only Poisson and background noise are considered, with the background noise model taken from \url{https://fermi.gsfc.nasa.gov/ssc/data/analysis/software/aux/iso_P8R3_SOURCE_V3_v1.txt}.}. We require $\mathrm{SNR}>5$ as a detection threshold. This criterion is conservative, as many LAT-detected GRB afterglows fall below this threshold. 

The opaque and dashed contours in Figure~\ref{E_n_plot} indicate the detectability for bursts placed at redshifts $z=0.2$ and $z=0.5$, respectively. At early times, the detectable parameter space expands as $0.1\nu_{\rm gap}$ shifts to lower energies within the LAT band during deceleration. At lower photon energies, a given energy fluence corresponds to a larger number of detected photons. Since $0.1\nu_{\rm gap}$ shifts to lower energies as $T_{\rm obs}^{-3/4}$, faster than the fluence declines (approximately as $T_{\rm obs}^{-1/2}$), the photon counting statistics around $0.1\nu_{\rm gap}$ improve with time during this phase. At later times, detectability becomes increasingly limited by background noise as exposure time grows, and the cutoff begins to shift outside the LAT sensitivity range. The optimal window for identifying the exponential cutoff is therefore typically a few hundred seconds after the burst. For $z=0.2$, there is substantial overlap between detectable events and those exhibiting clear spectral cutoff even at $\sim 3000$~s, while for $z=0.5$ the overlap is significantly reduced. The successful detection of a clear spectral cutoff relies on a GRB that \textbf{satisfies} several favorable conditions.

\section{Observations} \label{Obs}
Motivated by the results of the previous section, we focus on two well-studied GRB afterglows with early-time Fermi-LAT detections: GRB~190114C and GRB~130427A. We utilize multiwavelength data to model their early-time broadband spectra, adopting parameter ranges motivated by both PIC simulations and GRB observations. As discussed in Section~\ref{Sec_acc}, we fix $\varepsilon_e = 0.1$ and $p \simeq 2.4$, while allowing $10^{-4} \lesssim \varepsilon_B \leq 5 \times 10^{-3}$. For GRB observations, the isotropic-equivalent prompt energy $E_{\gamma,\mathrm{iso}}$ is typically well constrained. For small values of $\varepsilon_B$ ($\sim10^{-4}$ to $10^{-3}$), the radiative efficiency is inferred to be $E_{\gamma,\mathrm{iso}} / E_{\mathrm{iso}} \simeq 20\%$ \citep{2016MNRAS.461...51B}. We therefore adopt $E_{\mathrm{iso}} = 5 E_{\gamma,\mathrm{iso}}$ and treat the ambient density $n$ as a free parameter. 

\subsection{GRB 190114C} \label{Sec_190114C}
GRB~190114C was a bright long GRB with an isotropic-equivalent gamma-ray energy of $E_{\gamma,\mathrm{iso}} \simeq 3 \times 10^{53} \,\mathrm{erg}$ at $z = 0.42$~\citep{2020ApJ...890....9A}. Its relatively low redshift enabled extensive multiwavelength follow-up observations. Shortly after the trigger, the MAGIC Cherenkov telescopes detected VHE emission, marking the first unambiguous detection of TeV photons from a GRB and inaugurating ground-based VHE afterglow studies~\citep{MAGIC:2019lau}. 

\begin{figure} 
    \centering
    \includegraphics[width=\linewidth]{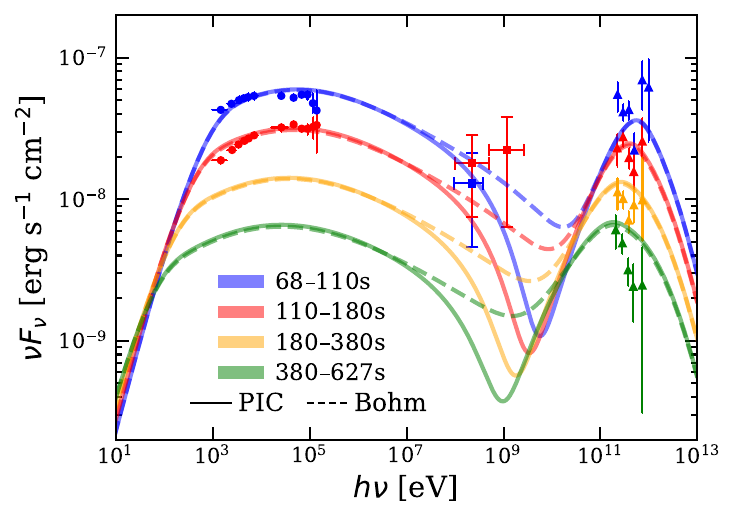}
    \vspace{-0.7cm}
    \caption{Spectral fitting of GRB~190114C from X-ray to TeV energies across multiple observation intervals. Different colors denote different time intervals, as indicated in the legend. Circles, squares, and triangles represent data from Swift/XRT–BAT, Fermi-LAT, and MAGIC, respectively, with error bars indicating $1\sigma$ uncertainties. Solid curves are based on the PIC acceleration model, while dashed curves correspond to the Bohm diffusion limit. Both of them adopt $\varepsilon_e = 0.1$, $\varepsilon_B = 10^{-3}$, $E_{\mathrm{iso}} = 1.5 \times 10^{54}\,\mathrm{erg}$, $n = 1\,\mathrm{cm^{-3}}$, and $p = 2.4$. }
    \label{190114C}
    \vspace{-0.3cm}
\end{figure}

Here we adopt the spectral energy distribution presented in \citet{MAGIC:2019irs} and focus on the first four MAGIC time intervals. The multiwavelength observations, including Swift/XRT–BAT and Fermi-LAT data, are summarized in Figure~\ref{190114C}, with different colors denoting the corresponding time intervals. The analyzed spectra span from $T_0$+68~s to $T_0$+625~s, where $T_0$ marks the onset of the prompt emission. The active prompt phase lasts for approximately 25~s, ensuring that the selected time window is minimally contaminated by prompt emission.

During the first time interval ($68–110$~s; blue points), the Swift/XRT–BAT observations indicate that the synchrotron component peaks in the X-ray band. At higher energies, up to $\lesssim 1$GeV, the spectrum declines with energy, consistent with the Fermi-LAT measurements. At even higher energies, the MAGIC detection above 0.2 TeV reveals a spectral hardening attributed to SSC emission. However, the locations of the spectral maximum and minimum remain poorly constrained owing to the limited coverage between the X-ray and TeV bands. A simultaneous broadband fit is therefore required to better constrain the maximum synchrotron frequency.

Within the parameter ranges discussed above, a reasonable fit to the spectrum in Figure~\ref{190114C} yields $\varepsilon_B = 10^{-3}$, $E_{\mathrm{iso}} = 1.5 \times 10^{54}\,\mathrm{erg}$, $n = 1\,\mathrm{cm^{-3}}$, and $p = 2.4$. The resulting model reproduces the observed spectrum over many orders of magnitude in photon energy and follows the observed temporal evolution, demonstrating the overall consistency of our framework. However, uncertainties in the Fermi-LAT measurements limit our ability to distinguish between the PIC acceleration model and the Bohm diffusion limit, as both yield similar spectral shapes for the same parameter set. The primary difference lies in the depth of the gap between the synchrotron and SSC components, but it is difficult to resolve owing to limited photon statistics at early times for this burst. A clearer detection of an exponential cutoff would likely be achievable if the LAT were not occulted by Earth and could maintain coverage up to $\sim T_0$+625~s~\citep{2020ApJ...890....9A}. Nevertheless, the presence of TeV observations plays a crucial role in reducing degeneracies in broadband spectral modeling.

\subsection{GRB 130427A} \label{Sec_130427A}
\begin{figure} 
    \centering
    \includegraphics[width=\linewidth]{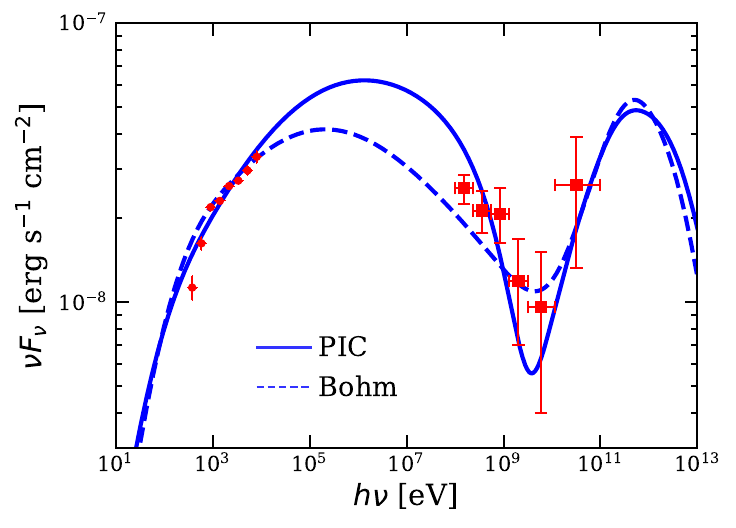}
    \vspace{-0.7cm}
    \caption{Spectral fitting of GRB~130427A from X-ray to GeV energies over the interval 138–750~s after the onset of the prompt emission. Circles and squares denote data from Swift/XRT and Fermi-LAT, respectively, with error bars indicating $1\sigma$ uncertainties. The solid curve shows the PIC-motivated acceleration model, while the dashed curve corresponds to the Bohm diffusion limit. Both models adopt $\varepsilon_e = 0.1$, $\varepsilon_B = 3 \times 10^{-4}$, and $E_{\mathrm{iso}} = 4 \times 10^{54}$~erg. The PIC model assumes $n = 1.5~\mathrm{cm^{-3}}$ and $p = 2.2$, whereas the Bohm model adopts $n = 1~\mathrm{cm^{-3}}$ and $p = 2.4$. }
    \label{130427A}
    \vspace{-0.3cm}
\end{figure}

GRB~130427A is the second brightest long GRB observed to date, with an isotropic-equivalent gamma-ray energy of $E_{\gamma,\mathrm{iso}} \simeq 8 \times 10^{53}\,\mathrm{erg}$ at a redshift of $z = 0.34$. A rich multiwavelength data set is available for this event, although it was not observed by MAGIC. Here we adopt the Fermi-LAT spectral energy distribution from \citet{2014Sci...343...42A} and focus on the interval from $T_0$+138~s to $T_0$+750~s, which provides sufficiently early coverage while maintaining good data quality. We adopt the Swift/XRT spectrum from \citet{2015ApJ...798...10R} for the same time interval\footnote{The spectral data in \citet{2015ApJ...798...10R} cover the interval from $T_0$+196~s to $T_0$+750~s. To extend the spectrum to earlier times ($T_0$+138 to $T_0$+196~s), we approximate the emission using the XRT light curve and photon index available from the Swift/XRT repository. We verify that the spectrum is dominated by emission from $T_0$+196 to $T_0$+750~s, ensuring that this approximation does not affect our results.}. All observational data are summarized in Figure~\ref{130427A}. The definition of $T_{90}$ for GRB~130427A is somewhat ambiguous, but the prompt emission is dominated by an initial pulse complex lasting approximately 18~s. Our selected time window is therefore not significantly contaminated by prompt emission.

Fortunately, Fermi-LAT continued to detect the burst up to $T_0$+750~s, allowing sufficient photon statistics to reveal spectral hardening associated with SSC emission. The minimum of the spectral energy distribution, $\nu_{\rm gap}$, occurs at energies of a few GeV, although its exact location is limited by photon statistics and background noise. We therefore perform a simultaneous broadband fit to fully exploit the available data and to place better constraints on $\nu_{\max}$.

A reasonable fit to the spectrum in Figure~\ref{130427A}, within the parameter ranges discussed above, yields $\varepsilon_B = 3 \times 10^{-4}$ and $E_{\mathrm{iso}} = 4 \times 10^{54}$~erg. For the PIC model, the best-fit parameters are $n = 1.5~\mathrm{cm^{-3}}$ and $p = 2.2$, while the Bohm limit favors $n = 1~\mathrm{cm^{-3}}$ and $p = 2.4$. Both acceleration prescriptions provide comparably good descriptions of the observed spectrum. Although the Fermi-LAT data for GRB~130427A provide improved constraints compared to GRB~190114C and suggest a possible spectral turnover at a few GeV, the uncertainties remain too large to discriminate between the PIC and Bohm acceleration scenarios.

Overall, while the data suggest a spectral drop followed by hardening, the uncertainties remain sufficiently large that even a single power-law spectrum in the Fermi-LAT band cannot be ruled out. An exponential cutoff could become detectable if the burst were closer by a factor of a few. Moreover, the lack of MeV band coverage introduces degeneracies in spectral modeling, highlighting the importance of future observations that bridge this energy range.

\section{Promising Probe of the Cutoff Location}
The burst properties most favorable for detecting a clear exponential cutoff, namely low circumburst densities and small kinetic energies, naturally point to short GRBs. Because sGRB afterglows are intrinsically faint, GeV detection generally requires a burst that is nearby, viewed on-axis, and observed within $\sim10^3$~s of the trigger. To date, only a few sGRBs have been detected by LAT, and none has data of sufficient quality for a meaningful spectral cutoff search.

As an illustrative example, we consider a GRB 170817-like event observed on-axis, but place it at a larger luminosity distance, increasing $d_{\rm L}$ from $40\, \mathrm{Mpc}$ to $200\, \mathrm{Mpc}$. The isotropic-equivalent energy is uncertain due to the dependence on jet-structure modeling, so we adopt $E_{\mathrm{iso}} = 2 \times 10^{52}$~erg, approximately the mean of the values reported in various studies summarized by \citet{2021ApJ...922..154M}. We take the remaining parameters directly from \citet{2021ApJ...922..154M}.

We then compute the time-averaged spectrum from $T_0$+100~s to $T_0$+600~s under the PIC prescription, together with simulated LAT data. The simulated data are generated using the same noise assumptions adopted in Section~\ref{parameter} for the burst detectability criterion, and the uncertainty is dominated by Poisson fluctuations in the source photon counts.

\begin{figure} 
    \centering
    \includegraphics[width=\linewidth]{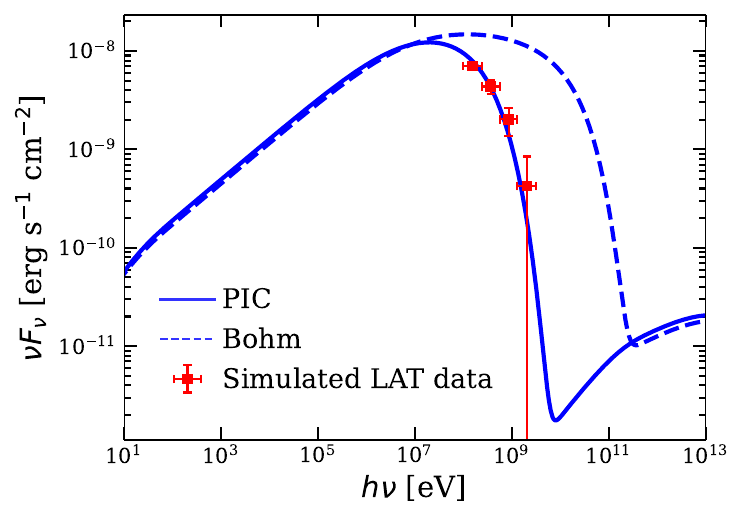}
    \vspace{-0.7cm}
    \caption{Simulated spectrum and Fermi-LAT data over 100--600 s after the onset of the prompt emission for a GRB 170817-like burst viewed on-axis at $d_{\rm L} = 200\, \mathrm{Mpc}$. Red squares show the simulated Fermi-LAT data with $1\sigma$ error bars. The solid and dashed curves correspond to the PIC-motivated acceleration model and the Bohm diffusion limit, respectively. Both models adopt $\varepsilon_e = 0.1$, $\varepsilon_B = 10^{-3}$, $E_{\mathrm{iso}} = 2 \times 10^{52}$~erg, $n = 10^{-3}~\mathrm{cm^{-3}}$, and $p = 2.17$. The spectral cutoffs are clearly visible, allowing the two acceleration models to be cleanly distinguished.}
    \label{170817}
    \vspace{-0.3cm}
\end{figure}

Figure~\ref{170817} shows a clear spectral cutoff at sub-GeV energies for the PIC model and at $\sim 10$~GeV for the Bohm limit, cleanly separating the two acceleration scenarios without the need for broadband spectral modeling. Although such an event generally requires a nearby burst viewed on-axis, the simulated data quality in this example is sufficiently high that these requirements could be somewhat relaxed while still preserving detectability of the cutoff. The estimated all-sky rate of detectable short GRBs within 200 Mpc is $1.3_{-0.8}^{+1.7} \, \mathrm{yr}^{-1}$~\citep{2020MNRAS.492.5011D}, and the parameters adopted in this example are broadly consistent with values inferred for short GRBs~\citep{2023ApJ...959...13R}. However, the rate of events that provide a promising probe of the cutoff location is likely lower, potentially by a factor of several or more, given the additional burst requirements (e.g., moderately high $E_{\mathrm{iso}}$) and the substantial uncertainties in the underlying population properties. Early LAT coverage is further limited by observational and technical constraints. Assuming nominal survey-mode operations and a standard LAT field of view of $2.4$~sr\footnote{The LAT effective area declines rapidly for boresight angles $\gtrsim 60^{\circ}$~\citep{2011MNRAS.416.3089B}, so the effective field of view is slightly smaller than the standard reported value.}, the expected detection rate for such events that remain within the field of view for $\sim 500$~s is $\lesssim 0.2\,{\rm yr^{-1}}$. The rate would increase if the required early-time coverage were reduced or LAT repointing were included. Therefore, detection of such an event in the near future remains feasible.

\section{Conclusion and Discussion} \label{discussion}
In this work, we investigate particle acceleration models in weakly magnetized relativistic collisionless shocks and apply them to GRB afterglows to place observational constraints on acceleration physics. We model the GRB afterglow spectrum and its temporal evolution during the relativistic deceleration phase. The primary observational diagnostic of the acceleration process is the synchrotron cutoff, which encodes the maximum attainable energy of accelerated electrons and thus the acceleration rate. A significant cutoff is expected only when the SSC component does not dominate near the maximum synchrotron frequency. We quantify the prominence of the cutoff using the flux ratio $\mathscr{F}$, with larger values corresponding to weaker SSC strength. Exploring a parameter space motivated by GRB observations and PIC simulations, we find that low-energy bursts in low-density environments exhibit a pronounced exponential cutoff at early times. The behavior is in good agreement with the theoretical prediction of $\mathscr{F}$ given in Equation~\eqref{theoretical}.

The conditions that need to be fulfilled for the detection of a clear exponential cutoff are not commonly satisfied. Low-energy bursts are only detectable at smaller distances, whereas early afterglow phases may be missed by Fermi-LAT owing to instrumental constraints. We examine two well-studied events, GRB~190114C and GRB~130427A, and find that the available LAT data are insufficient to robustly discriminate between PIC-motivated acceleration and the theoretical Bohm limit, while the broadband spectral modeling suffers from significant degeneracies. Future TeV observations with the Cherenkov Telescope Array Observatory \citep{2023arXiv230512888H} and MeV coverage from upcoming or proposed missions such as COSI \citep{2024icrc.confE.745T} and AMEGO-X \citep{2022JATIS...8d4003C} will be essential to break degeneracies. 

The burst properties favored for detecting a clear exponential cutoff, namely low circumburst densities and small kinetic energies, naturally point to short GRBs. Motivated by this, we present a simulated nearby, on-axis short GRB at $d_\mathrm{L} = 200$~Mpc as a promising case for clearly detecting the spectral cutoff location and distinguishing between different particle acceleration scenarios. For short GRBs within 200 Mpc with at least $\sim 500$~s of early LAT coverage, the estimated detection rate is $\sim 0.2\,{\rm yr^{-1}}$. While this should be regarded as a reasonable upper limit, the detection of such an event in the near future remains plausible.


GRB~221009A (the BOAT; \citealt{2023ApJ...946L..31B}) provides a notable counterexample. The burst saturated the LAT at early times, and once nominal operations resumed it remained within the field of view only briefly at $\sim120$~s, before re-entering at $\sim T_0+4000$~s \citep{2025ApJS..277...24A}, which has been outside our time window of interest. During this early valid interval, the photon index was already $<2$~\citep{2025ApJS..277...24A}, indicating that the LAT band was dominated by SSC emission and consistent with the event’s extreme energetics. The AGILE observations also support this interpretation~\citep{2023ApJ...956L..23T,2025A&A...701A..68B}. These considerations motivate excluding GRB~221009A from our main analysis, while its rapid SSC evolution may warrant a dedicated future study.

The PIC-simulation prescription for particle acceleation adopted here assumes a vanishingly small upstream magnetization. If this is not the case, the maximum particle energy may instead be limited by the influence of the large-scale ambient field, potentially shifting the synchrotron cutoff to lower energies. This is commonly referred to as the saturation limit, $\gamma_{\rm sat}$ (e.g., \citealt{2013ApJ...771...54S,2014MNRAS.439.2050R,2018MNRAS.477.5238P,Huang2022}). As shown in the Appendix~\ref{apped_3}, the parameter space and time range considered here remain safely outside this regime. 

Recent PIC simulations \citep{2024ApJ...963L..44G} reveal that the post-shock magnetic field is highly intermittent, with $\sim$1\% of the volume containing $\sim$50\% of the magnetic energy. This picture deviates from the commonly assumed uniform magnetic field strength throughout the shocked region and can affect both synchrotron and SSC emission. The long-term evolution of the magnetic field downstream of the shock may also play an important role. In addition, our analysis neglects the jet angular structure \citep[e.g.,][]{2021MNRAS.504..528J,2024ApJS..273...17W}, possible reprocessing of the ambient medium by gamma-ray photons~\citep{2002ApJ...565..808B,2016MNRAS.460.2036D,2022ApJ...933...74G,2025ApJ...986..211G}, and variations in the ambient density~\citep[e.g.,][]{2011MNRAS.418..583M}. Accounting for these effects represents important directions for future work.

\begin{acknowledgments}
We thank Zachary Davis for assistance with the \textsc{Tleco} code. Z. W. and D. Giannios acknowledge support from the NSF AST-2308090 and AST-2510569 grants. PB is supported by a grant from the National Aeronautics and Space Administration (NASA 80NSSC24K0770), by a grant (no. 2024788) from the United States-Israel Binational Science Foundation (BSF), Jerusalem, Israel and by a grant (no. 1649/23) from the Israel Science Foundation. D. Grošelj is supported by the Research Foundation—Flanders (FWO) Senior Postdoctoral Fellowship 12B1424N. L.S. acknowledges support from DoE Early Career Award DE-SC0023015, NASA ATP 80NSSC24K1238, NASA ATP 80NSSC24K1826, and NSF AST-2307202. This research was facilitated by the Multimessenger Plasma Physics Center (MPPC) NSF grant PHY-2206609 to L.S., and by a grant from the Simons Foundation (MP-SCMPS-0000147, to L.S.). This research has made use of NASA's Astrophysics Data System Bibliographic Services.
\end{acknowledgments}

%
\facilities{Fermi(LAT), Swift(XRT and BAT), MAGIC}

\software{\textsc{SciPy} \citep{2020NatMe..17..261V}, \textsc{NumPy} \citep{2020Natur.585..357H}, \textsc{Matplotlib} \citep{2007CSE.....9...90H}, \textsc{Tleco} \citep{2024ApJ...976..182D}}


\appendix

\section{Characteristic Breaks and Analytical Spectra} \label{apped_1}
In this section, we derive the characteristic frequencies of the afterglow spectrum and their dependence on the burst parameters. All critical Lorentz factors of the electron energy distribution are unprimed and defined in the shocked fluid frame, while all characteristic frequencies are given in the observer frame with $z=0$.

Using the Blandford–McKee self-similar solution \citep{1976PhFl...19.1130B} and the definition of the observer timescale in Equation~\eqref{timescales}, the bulk Lorentz factor of the shocked fluid can be written as
\begin{equation}
    \Gamma \simeq 200 \, E_{{\mathrm{iso}},54}^{~1/8} \, n_0^{-1/8} \, T_{\mathrm{obs}, 2}^{-3/8}
\end{equation}
The minimum electron Lorentz factor given in Equation~\eqref{gamma_min} then becomes
\begin{equation}
    \gamma_{\min } \simeq 3.5 \times 10^4 \left(\frac{p-2}{p-1}\right) \,\varepsilon_{\mathrm{e},-1} \,E_{{\mathrm{iso}},54}^{~1/8} \,n_0^{-1/8}\, T_{\mathrm{obs}, 2}^{-3/8} 
\end{equation}
where $\varepsilon_{\mathrm{e}} = 0.1,\varepsilon_{\mathrm{e},-1}$. The corresponding characteristic synchrotron frequency is
\begin{equation}
    h \nu_{\rm m} \simeq 10^4 \left(\frac{p-2}{p-1}\right)^2 \,\varepsilon_{\mathrm{e},-1}^{~2} \,\varepsilon_{B,-2.5}^{~1/2} \,E_{{\mathrm{iso}},54}^{~1/2} \,T_{\mathrm{obs}, 2}^{-3/2} \mathrm{~eV}
\end{equation}
Another important Lorentz factor is the synchrotron cooling Lorentz factor $\gamma^{\rm syn}_{\rm c}$, defined by equating the synchrotron cooling timescale to the effective cooling timescale, $\gamma^{ \rm syn }_{\rm c} = \dot{\gamma}_{\text {c }}^{\prime \, \rm syn} t_{\rm cool}^{\prime}$. The effective cooling timescale accounts for both particle escape ($t'_{\rm esc}$) and adiabatic expansion ($t'_{\rm ad}$) timescales. Following \citet{2026MNRAS.546ag101A}, the adiabatic timescale is 
\begin{equation}
\frac{1}{t_{\mathrm{ad}}^{\prime}}=\frac{3 \Gamma \beta c}{r}\left(1-\frac{1}{3} \frac{d \log \Gamma}{d \log r}\right) = \frac{9 \Gamma c}{2r}.
\end{equation}
Together with the escape time $t'_{\rm esc}$ (Equation~\eqref{t_esc}), the effective cooling timescale is
\begin{equation}
    t'_{\rm cool} = \left(\frac{1}{t'_{\rm esc}} + \frac{1}{t_{\mathrm{ad}}^{\prime}}\right)^{-1} = \frac{6r}{59\Gamma c} \simeq 0.1 \frac{r}{\Gamma c}.
\end{equation}
The synchrotron cooling Lorentz factor is then
\begin{equation}
    \gamma^{\rm syn}_{\rm c} = \frac{6\pi  m_{\rm e} c}{ \sigma_\mathrm{T} B'^2 t'_{\rm cool}} = 4\times10^4 \,\varepsilon_{B,-2.5}^{-1} \,E_{{\mathrm{iso}},54}^{-3/8}\, n_0^{-5/8} \,T_{\mathrm{obs}, 2}^{~1/8} 
\end{equation}
with the associated cooling frequency
\begin{equation}
    h \nu^{\rm syn}_\mathrm{c} = 1.3 \times 10^4\, \varepsilon_{B, -2.5}^{-3 / 2} \,E_{{\mathrm{iso}},54}^{-1 / 2} \,n_{0}^{-1} \,T_{\mathrm{obs}, 2}^{-1 / 2} \mathrm{~eV}
\end{equation}
For most of the parameter space considered here, Klein–Nishina effects cannot be neglected, and the inverse-Compton cooling rate becomes energy dependent. As a result, the Compton $Y$ parameter depends on the energy of the scattered electrons. The total cooling break should include the contribution from SSC losses and is determined by
\begin{equation} \label{gamma_c_tot}
\gamma_\mathrm{c}\left[1+Y\left(\gamma_\mathrm{c}\right)\right]=\gamma_\mathrm{c}^{\text {syn }}, \nu_\mathrm{c} \left[1+Y\left(\gamma_\mathrm{c}\right)\right]^2= \nu^{\rm syn}_\mathrm{c},
\end{equation}
where $Y(\gamma_{\rm c})$ is the Compton parameter evaluated at $\gamma_{\rm c}$. $Y(\gamma_{\rm c})$ can be obtained iteratively following \citet{2024A&A...690A.281P}, or estimated using the approach of \citet{2009ApJ...703..675N} with details provided in Appendix~\ref{apped_2}. For the parameter space relevant to this work, $Y(\gamma_{\rm c})$ is modest or low ($Y\left(\gamma_\mathrm{c}\right) \lesssim 1$), so that $\nu_{\rm c}>\nu_{\rm m}$ remains satisfied. In this regime, the synchrotron $\nu F^{\,\rm syn}_\nu$ spectrum peaks at $\nu_{\rm c}$, while the peak of the $F^{\,\rm syn}_\nu$ spectrum occurs at $\nu_{\rm m}$ and can be estimated following \citet{1998ApJ...497L..17S},
\begin{equation}
F^{\,\rm syn}_{\rm \nu, peak  } = \frac{\Gamma N P'_{\rm \nu', peak}}{4 \pi D_{\rm L}^2} 
\end{equation}
where $D_{\rm L}$ is the luminosity distance, $N$ is the total number of particles swept by the equivalent spherical blast wave, and the peak spectral power per electron in the comoving frame is
\begin{equation}
    P'_{\rm \nu', peak} = \frac{\sqrt{3} \mathrm{e}^3 B'}{m_\mathrm{e} c^2}.
\end{equation}
Following \citet{2009ApJ...703..675N} and \citet{2024A&A...690A.281P}, we approximate the inverse-Compton cross section by assuming that scattering is fully suppressed in the KN regime. Under this assumption, a synchrotron photon emitted by an electron of Lorentz factor $\gamma$ can be upscattered only by electrons with Lorentz factors below a critical value $\widehat{\gamma}$, such that scattering remains in the Thomson regime,
\begin{equation} \label{gamma_hat}
    \widehat{\gamma}=\frac{w_{\mathrm{KN}} \Gamma m_{\mathrm{e}} c^2}{h \nu (\gamma)}.
\end{equation}
where we adopt $w_{\rm KN}=0.2$ following \citet{2022MNRAS.512.2142Y}. The corresponding characteristic KN breaks are
\begin{equation}
    \widehat{\gamma_{\rm m}} = \frac{w_{\mathrm{KN}} \Gamma m_{\mathrm{e}} c^2}{h \nu_\mathrm{m}}  = 2 \times 10^3 \left(\frac{p-1}{p-2}\right)^2 \, \varepsilon_{\mathrm{e},-1}^{-2} \, \varepsilon_{B,-2.5}^{-1/2} \, E_{{\mathrm{iso}},54}^{-3/8} \, n_0^{-1/8} \, T_{\mathrm{obs}, 2}^{~9/8},
\end{equation}
and
\begin{equation} \label{gamma_c_hat}
    \widehat{\gamma_{\rm c}} = \frac{w_{\mathrm{KN}} \Gamma m_{\mathrm{e}} c^2}{h \nu_\mathrm{c}} \left[1+Y\left(\gamma_\mathrm{c}\right)\right]^{2} =1.6 \times 10^3 \left[1+Y\left(\gamma_\mathrm{c}\right)\right]^{2} \, \varepsilon_{B, -2.5}^{~3 / 2} \, E_{{\mathrm{iso}},54}^{~5 / 8} \, n_{0}^{7/8} \, T_{\mathrm{obs}, 2}^{~1 / 8}.
\end{equation}
Given these characteristic breaks, the SSC spectrum can be constructed following the prescription of \citet{2024A&A...690A.281P}. The synchrotron and SSC spectra shown in Figure~\ref{fig:illustration} correspond to Case~S4 in \citet{2024A&A...690A.281P}\footnote{The spectral slope between $\gamma_{\rm m}$ and $\gamma_{\rm c}$ should be $p$, and the value quoted in the original reference is a typo.}.

\section{Derivation and Discussion of $\mathscr{F}$} \label{apped_2}
Here we derive the analytical approximation for $\mathscr{F}$ given in Equation~\eqref{theoretical}. As discussed in Section~\ref{parameter}, values of $\mathscr{F}\gtrsim10$ cannot be produced by a pure synchrotron power-law spectrum and therefore require a contribution from the synchrotron cutoff. This implies that the spectral minimum $\nu_{\rm gap}$ must be greater than the cutoff frequency $\nu_{\max}$. If $0.1\nu_{\rm gap}>\nu_{\max}$, $\mathscr{F}$ would probe only the exponential decay and become very large. We neglect this case, as the SSC contribution is negligible and the cutoff is always clear. We therefore focus on the regime  $0.1\nu_{\rm gap} < \nu_{\max} < \nu_{\rm gap}$.

In this regime, the power-law decline of the synchrotron spectrum is subdominant compared to the exponential cutoff, and the flux ratio is dominated by the cutoff contribution. We thus approximate 
\begin{equation}
    \nu F_\nu (0.1 \nu_{\rm gap}) \approx \nu F_{\nu} (\nu_{\max}) \approx \nu F^{\,\rm syn}_\nu(\nu_{\max}). 
\end{equation}
using the fact that the emission at $\nu_{\max}$ is synchrotron dominated since $\nu_{\max}<\nu_{\rm gap}$. 

At $\nu_{\rm gap}$, the synchrotron and SSC components become comparable, so that
\begin{equation}
    \nu F_\nu (\nu_{\rm gap}) \approx \nu F^{\,\rm IC}_\nu (\nu_{\rm gap})
\end{equation}
Moreover, because the power-law rise in the SSC spectrum is insignificant compared to the exponential drop, we can further approximate 
\begin{equation}
    \nu F^{\,\rm IC}_\nu (\nu_{\rm gap}) \approx \nu F^{\,\rm IC}_\nu (\nu_{\max})
\end{equation}
Combining these relations, we arrive at the following approximation for $\mathscr{F}$:
\begin{equation} \label{F_app}
    \mathscr{F} = \frac{\nu F_\nu (0.1 \nu_{\rm gap})}{\nu F_\nu (\nu_{\rm gap})} \approx \frac{\nu F^{\,\rm syn}_\nu (\nu_{\max})}{\nu F^{\rm IC}_\nu (\nu_{\max})}.
\end{equation}
With these approximations, $\mathscr{F}$ reduces to the ratio of the synchrotron and SSC amplitudes of $\nu F_\nu$ evaluated at $\nu_{\max}$. To estimate the relative normalization of the two components, the Compton parameter naturally enters. In particular,
\begin{equation}
    Y\left(\gamma_\mathrm{c}\right) = \frac{\nu F_{\nu} (\nu^{\rm IC}_{\rm peak})}{\nu F_{\nu} (\nu_{\rm c})}\simeq \frac{\nu F^{\rm IC}_{\nu} (\nu^{\rm IC}_{\rm peak})}{\nu F^{\,\rm syn}_\nu (\nu_{\rm c})}
\end{equation}
where we have used the fact that, for the parameter space of interest, the synchrotron peak occurs at $\nu_{\rm peak}^{\rm syn}=\nu_{\rm c}$. Following \citet{2009ApJ...703..675N}, the Compton parameter in the slow-cooling regime satisfies
\begin{equation} \label{Y_c_general}
Y\left(\gamma_\mathrm{c}\right)\left[1+Y\left(\gamma_\mathrm{c}\right)\right] \approx \frac{1}3{}\frac{\varepsilon_{\mathrm{e}}}{\varepsilon_{\rm B}}\left(\frac{\gamma_{\rm c}}{\gamma_{\rm m}}\right)^{2-p}\left(\frac{\min \left\{\gamma_{\rm c}, \widehat{\gamma}_{\rm c}\right\}}{\gamma_{\rm c}}\right)^{\frac{3-p}{2}}.
\end{equation}
The prefactor $1/3$ accounts for the fact that the borders of the shock fluid expand at a speed of $c/3$ rather than $c$. Using the expression for ${\gamma}_{\rm c}$ and $\widehat{\gamma}_{\rm c}$ derived in Equation~\eqref{gamma_c_tot} and Equation~\eqref{gamma_c_hat}, we obtain
\begin{equation}
    \frac{\widehat{\gamma_{\rm c}}}{\gamma_{\rm c}} = 0.04 \,\left[1+Y\left(\gamma_\mathrm{c}\right)\right]^3\, \varepsilon_{B, -2.5}^{5 / 2}\, E_{{\mathrm{iso}},54}\, n_{0}^{3/2}.
\end{equation}
Since a clear synchrotron cutoff requires $Y(\gamma_{\rm c})\lesssim\mathcal{O}(1)$, we typically have $\gamma_{\rm c}>\widehat{\gamma}_{\rm c}$ in the regime of interest, where the KN suppression becomes important. Then Equation~(\ref{Y_c_general}) reduces to
\begin{equation} \label{Y_c}
\begin{aligned}
   Y &\left(\gamma_\mathrm{c}\right) \approx \frac{1}{3}\frac{\varepsilon_{\mathrm{e}}}{\varepsilon_{\rm B}}\left(\frac{\gamma_{\rm c}}{\gamma_{\rm m}}\right)^{2-p}\left(\frac{\widehat{\gamma}_{\rm c}}{\gamma_{\rm c}}\right)^{\frac{3-p}{2}} \left[1+Y\left(\gamma_\mathrm{c}\right)\right] ^{-1}\\
   & =13.7~\left(\frac{p-2}{p-1}\right)^{p-2} \left[1+Y\left(\gamma_\mathrm{c}\right)\right]^{\frac{3-p}{2}} \, \varepsilon_{\mathrm{e},-1}^{~p-1} \, \varepsilon_{B, -2.5}^{\frac{3-p}{4}} \, E_{{\mathrm{iso}},54}^{~\frac{1}{2}} \, n_{0}^{\frac{5-p}{4}} \, T_{\mathrm{obs}, 2}^{-\frac{p-2}{2}}.
\end{aligned}
\end{equation}
For $p\simeq2.4$, the strongest dependence is on the circumburst density, and requiring $Y(\gamma_{\rm c})\lesssim\mathcal{O}(1)$ implies $n\lesssim1\,\mathrm{cm^{-3}}$, consistent with the condition $\gamma_{\rm c}>\widehat{\gamma}_{\rm c}$ assumed above.

Using $Y(\gamma_{\rm c})$, we estimate the ratio between $\nu F_\nu^{\,\rm syn}(\nu_{\max})$ and $\nu F_\nu^{\rm IC}(\nu_{\max})$. For $\gamma_{\rm c}>\widehat{\gamma}_{\rm c}$ and $Y(\gamma_{\rm c})\lesssim\mathcal{O}(1)$, the synchrotron spectrum between $\nu_{\rm c}$ and $\nu_{\max}$ follows the standard power-law decline, $\nu F_\nu\propto\nu^{(2-p)/2}$ \citep{2009ApJ...703..675N,2024A&A...690A.281P}. Therefore,
\begin{equation} \label{syn_maxx}
\begin{aligned}
    \nu & F^{\,\rm syn}_\nu (\nu_{\max})  = \left( \frac{\nu_{\max}}{\nu_\mathrm{c}} \right)^{\frac{2-p}{2}}\nu F^{\,\rm syn}_\nu (\nu_{\rm c}) \\
    &=\left( \frac{\nu_{\max}}{\nu_\mathrm{c}} \right)^{\frac{2-p}{2}}  \frac{\nu F^{\rm IC}_{\nu} (\nu^{\rm IC}_{\rm peak})}{Y\left(\gamma_\mathrm{c}\right)},
\end{aligned}
\end{equation}
where the second equality follows from the definition of the Compton parameter. To relate $\nu F_\nu^{\rm IC}(\nu_{\rm peak}^{\rm IC})$ to $\nu F_\nu^{\rm IC}(\nu_{\max})$, we next derive the SSC characteristic frequencies in between. The SSC peak frequency is given by \citep{2009ApJ...703..675N}
\begin{equation}
    h \nu^{\rm IC}_{\rm peak} = \frac{4}{3}\gamma_{\rm c} \widehat{\gamma_{\rm c}} \nu_{\rm c} = 1.1 \times 10^3 \left[1+Y\left(\gamma_\mathrm{c}\right)\right]^{-1} \varepsilon_{B,-2.5}^{-1} \, E_{{\mathrm{iso}},54}^{-1 / 4} \, n_{0}^{-3 / 4} \, T_{\mathrm{obs}, 2}^{-1 / 4} \mathrm{~GeV},
\end{equation}
while another important SSC break is
\begin{equation}
    h \nu_\mathrm{m}^{\rm IC} = \frac{4}{3}\gamma_\mathrm{m}^2 \nu_\mathrm{m}  = 1.6 \times 10^4 \left(\frac{p-2}{p-1}\right)^4 \varepsilon_{\mathrm{e},-1}^{~4} \, \varepsilon_{B,-2.5}^{~1/2} \, E_{{\mathrm{iso}},54}^{~3/4} \, T_{\mathrm{obs}, 2}^{-9/4}
    \mathrm{~GeV}.
\end{equation}
For fiducial parameters and $p\simeq2.4$, we find that $\nu_{\rm m}^{\rm IC}\simeq100\,\mathrm{GeV}<\nu_{\rm peak}^{\rm IC}$. Moreover, $\nu_{\rm m}^{\rm IC}$ decreases rapidly with observer time, so the ordering $\nu_{\rm m}^{\rm IC}<\nu_{\rm peak}^{\rm IC}$ is expected throughout the parameter space and time range of interest. Between $\nu_{\rm m}^{\rm IC}$ and $\nu_{\rm peak}^{\rm IC}$, the SSC spectrum is relatively flat, with $\nu F_\nu^{\rm IC}\propto\nu^{(3-p)/2}$, yielding
\begin{equation} \label{v_m_f}
    \nu F^{\rm IC}_{\nu} (\nu^{\rm IC}_{\rm m}) = \nu F^{\rm IC}_{\nu} (\nu^{\rm IC}_{\rm peak}) \left( \frac{\nu_m^{\rm IC}}{\nu^{\rm IC}_{\rm peak}}\right)^{\frac{3-p}{2}}.
\end{equation}
From this point onward, we adopt the PIC prescription for the maximum synchrotron frequency. At observer times of a few hundred seconds, the SSC break remains above the synchrotron cutoff, 
\begin{equation} 
    h \nu_{\mathrm{max,PIC}} \simeq 0.8 \, \varepsilon_{B,-2.5}^{-1 / 6} \, E_{{\mathrm{iso}},54}^{~1 / 4} \, n_{0}^{-1 / 12}  \,T_{\mathrm{obs}, 2}^{-3 / 4} \mathrm{~GeV}.
\end{equation}
Below $\nu_{\rm m}^{\rm IC}$, the SSC spectrum rises more steeply, following $\nu F_\nu^{\rm IC}\propto\nu^{4/3}$. Consequently,
\begin{equation} \label{v_max_f}
    \nu F^{\rm IC}_\nu (\nu_{\rm max, PIC}) = \nu F^{\rm IC}_{\nu} (\nu^{\rm IC}_{\rm m}) \left(\frac{\nu_{\rm max, PIC}}{\nu^{\rm IC}_{\rm m}}\right)^{\frac{4}{3}}.
\end{equation}
Between $\nu_{\rm m}^{\rm IC}$ and $\nu_{\rm peak}^{\rm IC}$ the SSC spectrum is relatively flat, whereas below $\nu_{\rm m}^{\rm IC}$ it rises much more steeply. As a result, even a modest separation between $\nu_{\rm max,PIC}$ and $\nu_{\rm m}^{\rm IC}$ substantially suppresses $\nu F_\nu^{\rm IC}(\nu_{\rm max,PIC})$ relative to its peak value, making the SSC component subdominant at $\nu_{\rm max,PIC}$ and enhancing the visibility of the synchrotron cutoff. Because $\nu_{\rm m}^{\rm IC}\propto T_{\rm obs}^{-9/4}$ decreases rapidly with time, it quickly approaches $\nu_{\rm max,PIC}$ and can even fall below it at $T_{\rm obs}\gtrsim10^3$~s. This behavior explains why a clear synchrotron cutoff is strongly favored at early observer times.

If $\nu_{\rm max,PIC}<\nu_{\rm m}^{\rm IC}$, then using Equations~\eqref{syn_maxx}, \eqref{v_m_f}, and \eqref{v_max_f}, we obtain
\begin{equation}
\begin{aligned}
    & \frac{\nu F^{\,\rm syn}_\nu (\nu_{\rm max, PIC})}{\nu F^{\rm IC}_\nu (\nu_{\rm max, PIC})} =\left( \frac{\nu_{\rm max, PIC}}{\nu_\mathrm{c}} \right)^{\frac{2-p}{2}}  \frac{\nu F^{\rm IC}_{\nu} (\nu^{\rm IC}_{\rm peak})}{\nu F^{\rm IC}_\nu (\nu_{\rm max, PIC})Y\left(\gamma_\mathrm{c}\right)} \\
    & ={Y\left(\gamma_\mathrm{c}\right)}^{-1} \left( \frac{\nu_{\rm max, PIC}}{\nu_\mathrm{c}} \right)^{\frac{2-p}{2}}  \left( \frac{\nu_m^{\rm IC}}{\nu^{\rm IC}_{\rm peak}}\right)^{\frac{p-3}{2}} \left(\frac{\nu_{\rm max, PIC}}{\nu^{\rm IC}_{\rm m}}\right)^{-\frac{4}{3}}. 
\end{aligned}
\end{equation}
Combining the above expression with Equations~\eqref{F_app} and \eqref{Y_c}, the flux ratio can be written as
\begin{equation}
\begin{aligned}
    &\mathscr{F} \approx \frac{\nu F^{\,\rm syn}_\nu (\nu_{\rm max, PIC})}{\nu F^{\rm IC}_\nu (\nu_{\rm max, PIC})} \\
    &={Y\left(\gamma_\mathrm{c}\right)}^{-1} \left( \frac{\nu_{\rm max, PIC}}{\nu_\mathrm{c}} \right)^{\frac{2-p}{2}}  \left( \frac{\nu_m^{\rm IC}}{\nu^{\rm IC}_{\rm peak}}\right)^{\frac{p-3}{2}} \left(\frac{\nu_{\rm max, PIC}}{\nu^{\rm IC}_{\rm m}}\right)^{-\frac{4}{3}} \\
    &= 4 \times 10^7 (4\times 10^{3})^{-\frac{p}{2}} \left(\frac{p-2}{p-1}\right)^{\frac{4}{3} + p}\left[1 + Y\left(\gamma_\mathrm{c}\right) \right]^{-1} \varepsilon_{\mathrm{e},-1}^{p+\frac{1}{3}} \, \varepsilon_{B, -2.5}^{\frac{3p -7}{9}} \, E_{{\mathrm{iso}},54}^{\frac{p}{8} - \frac{7}{12}} \, n_{0}^{\frac{p}{6} - \frac{97}{72}} \, T_{\mathrm{obs}, 2}^{-\frac{3p+2}{8}}.
\end{aligned}
\end{equation}
Equation~\eqref{Y_c} can be used for a quick estimate of $Y(\gamma_{\rm c})$. For $Y(\gamma_{\rm c})<1$, we have $(1+Y)\simeq1$, while for $Y(\gamma_{\rm c})\gtrsim1$, $(1+Y\left(\gamma_\mathrm{c}\right)) \sim Y\left(\gamma_\mathrm{c}\right)$ while the Compton parameter is still expected to be of order unity. In this latter case, we may further approximate $\left[1+Y(\gamma_{\rm c})\right]^{(3-p)/2}\simeq1$ since the exponent $(3-p)/2$ is small for $p \simeq 2.4$. The resulting expressions for $\mathscr{F}$ are therefore
\begin{equation}
\mathscr{F} \simeq
\begin{cases}
{4 \times 10^7 (4\times 10^{3})^{-\frac{p}{2}} \left(\frac{p-2}{p-1}\right)^{\frac{4}{3} + p} \varepsilon_{\mathrm{e},-1}^{~p+\frac{1}{3}} \,  \varepsilon_{B, -2.5}^{\frac{3p -7}{9}} \, E_{{\mathrm{iso}},54}^{\frac{p}{8} - \frac{7}{12}} \, n_{0}^{\frac{p}{6} - \frac{97}{72}} \, T_{\mathrm{obs}, 2}^{-\frac{3p}{8} - \frac{1}{4}}} & \text { for } Y\left(\gamma_\mathrm{c}\right) <1 \\ 
{3 \times 10^6 (4\times 10^{3})^{-\frac{p}{2}} \left(\frac{p-2}{p-1}\right)^{\frac{10}{3}} \varepsilon_{\mathrm{e},-1}^{~\frac{4}{3}} \, \varepsilon_{B, -2.5}^{\frac{7p}{12} - \frac{3}{2}} \, E_{{\mathrm{iso}},54}^{\frac{p}{8} - \frac{13}{12}} \, n_{0}^{\frac{5p}{12} - \frac{187}{72}} \, T_{\mathrm{obs}, 2}^{~\frac{p}{8} - \frac{5}{4}}} & \text { for } Y\left(\gamma_\mathrm{c}\right) \gtrsim 1.
\end{cases}
\end{equation}
These expressions show that $\mathscr{F}$ decreases with increasing burst energy $E_{\rm iso}$, circumburst density $n$, and observer time $T_\mathrm{obs}$, with only a weak dependence on the electron spectral index $p$. For $p\simeq2.4$, the expressions reduce to 
\begin{equation} 
\mathscr{F} \simeq
\begin{cases}
{17.7 \,\varepsilon_{\mathrm{e},-1}^{~2.7} \,\varepsilon_{B, -2.5}^{~0.02} \,E_{{\mathrm{iso}},54}^{-0.28} \,n_{0}^{-0.95} \,T_{\mathrm{obs}, 2}^{-1.15}} & \text { for } Y\left(\gamma_\mathrm{c}\right) < 1 \\ 
{2.2 \,\varepsilon_{\mathrm{e},-1}^{~1.3} \,\varepsilon_{B, -2.5}^{-0.1} \,E_{{\mathrm{iso}},54}^{-0.78} \,n_{0}^{-1.6} \,T_{\mathrm{obs}, 2}^{-0.95}} & \text { for } Y\left(\gamma_\mathrm{c}\right) \gtrsim 1.
\end{cases}
\end{equation}
A large value of $\mathscr{F}$, corresponding to a more pronounced synchrotron cutoff, is therefore favored for low-energy explosions in low-density environments at early times. Increasing $n$ raises the number of relativistic electrons and therefore the optical depth of synchrotron photons, enhancing SSC emission and reducing $\mathscr{F}$. Moreover, larger $E_{\rm iso}$ and $n$ shift $\nu_{\rm c}$ to lower frequencies, weakening KN suppression and increasing the Compton $Y$ parameter, which further boosts SSC relative to synchrotron emission and suppresses $\mathscr{F}$.

At $T_{\rm obs}\sim10^3$~s, we typically have $\nu_{\rm max,PIC}>\nu_{\rm m}^{\rm IC}$, and a similar procedure can be followed, yielding
\begin{equation}
\mathscr{F} \simeq
\begin{cases}
{2.3 \times 10^8 (1.2\times 10^{-8})^{\frac{p}{2}} \left(\frac{p-2}{p-1}\right)^{2 - p} \varepsilon_{\mathrm{e},-1}^{1-p} \, \varepsilon_{B, -2.5}^{-\frac{2}{3}} \, E_{{\mathrm{iso}},54}^{-\frac{p}{8} - \frac{1}{2}} \, n_{0}^{\frac{p}{8} - \frac{4}{3}} \, T_{\mathrm{obs}, 2}^{\frac{3p}{8} - \frac{1}{2}}} & \text { for } Y\left(\gamma_\mathrm{c}\right) < 1 \\ 
{1.7 \times 10^7 (1.2\times 10^{-8})^{\frac{p}{2}} \left(\frac{p-2}{p-1}\right)^{4 -2p} \varepsilon_{\mathrm{e},-1}^{2 - 2p} \, \varepsilon_{B, -2.5}^{\frac{p}{4} - \frac{17}{12}} \, E_{{\mathrm{iso}},54}^{-\frac{p}{8} - 1} \, n_{0}^{\frac{3p}{8} - \frac{31}{12}} \, T_{\mathrm{obs}, 2}^{\frac{7p}{8} - \frac{3}{2}}} & \text { for } Y\left(\gamma_\mathrm{c}\right) \gtrsim 1.
\end{cases}
\end{equation}
For $p\simeq2.4$, the expressions reduce to
\begin{equation} 
\mathscr{F} \simeq
\begin{cases}
{0.12 \,\varepsilon_{\mathrm{e},-1}^{-1.4} \, \varepsilon_{B, -2.5}^{-0.7} \,E_{{\mathrm{iso}},54}^{-0.8} \,n_{0}^{-1} \,T_{\mathrm{obs}, 2}^{~0.4}} & \text { for } Y\left(\gamma_\mathrm{c}\right) < 1 \\ 
{0.014 \,\varepsilon_{\mathrm{e},-1}^{-2.8} \, \varepsilon_{B, -2.5}^{-0.8} \,E_{{\mathrm{iso}},54}^{-1.3} \,n_{0}^{-1.7} \,T_{\mathrm{obs}, 2}^{~0.6}} & \text { for } Y\left(\gamma_\mathrm{c}\right) \gtrsim 1.
\end{cases}
\end{equation}
In this regime, $\mathscr{F}$ shows a modest increase with observer time. However, this behavior has limited practical relevance, as the time dependence is weak and the absolute values of $\mathscr{F}$ are already substantially smaller than those at $T_{\rm obs}\sim100$~s. Moreover, at late times the source fades and background contamination becomes more important, while the spectral cutoff progressively shifts outside the LAT sensitivity range, consistent with the shrinking detectability region in Figure~\ref{E_n_plot}.

\section{The Saturation Limit} \label{apped_3}
Here we evaluate the magnetization-limited maximum particle energy under realistic physical conditions and compare it with the cooling-limited value. The saturation limit set by the upstream magnetization is given by~\citet{2013ApJ...771...54S},
\begin{equation}
\gamma_{\mathrm{sat}} \simeq 2 \frac{ m_{\rm p}}{m_{\rm e}}\Gamma \sigma^{-1 / 4},
\end{equation}
where $\sigma$ is the magnetization of the upstream plasma. For a typical ISM magnetic field, $B_{\mathrm{ISM}} \equiv 3 B_{\mathrm{ISM},-5.5} \mu \mathrm{G}$, the corresponding magnetization is
\begin{equation}
    \sigma=\frac{B_{\mathrm{ISM}}^2}{4 \pi n m_p c^2} \simeq 0.5 \times 10^{-9} B_{\mathrm{ISM},-5.5}^2 n_0^{-1}.
\end{equation}
The ratio between the cooling-limited $\gamma_{\rm max}$ and the saturation limit is then
\begin{equation}
    \frac{\gamma_\mathrm{max, PIC}}{\gamma_\mathrm{sat}} = 0.087 \, \varepsilon_{B, -2.5}^{-1/3}  \, E_{{\mathrm{iso}},54}^{-1/8} \, n_0^{-1/24} \, T_{\mathrm{obs}, 2}^{~3/8} \, \sigma^{1/4}_{-9},
\end{equation}
which is small and only weakly dependent on the burst parameters. It remains below unity even for $\varepsilon_{B} \simeq 10^{-4}$ up to $T_{\mathrm{obs}} = 3000$~s, so the saturation limit can be safely neglected in this study. In a stellar wind environment, if the magnetic field in the wind is primarily toroidal, the magnetization remains approximately constant with radius. For a typical Wolf–Rayet progenitor, one may adopt $\sigma \simeq 1.7 \times 10^{-8}$~\citep{1993ApJ...402..271E,2013ApJ...771...54S}, although the wind magnetization is currently poorly constrained and could in principle be much higher. For $\sigma \sim 10^{-8}$,
\begin{equation}
    \frac{\gamma^{\rm wind}_\mathrm{max, PIC}}{\gamma^{\rm wind}_\mathrm{sat}} = 0.14 \, \varepsilon_{B, -2.5}^{-1/3}  \, E_{{\mathrm{iso}},54}^{-1/12} \, A_{*}^{-1/12} \, T_{\mathrm{obs}, 2}^{~5/12} \, \sigma^{1/4}_{-8}.
\end{equation}
Therefore, the saturation limit is relevant on minute-to-hour timescales only for strongly magnetized winds or sufficiently small $\varepsilon_{B}$. Generally, the saturation limit becomes important only at late times~\citep{2026arXiv260121827H}. 

\bibliography{references}{}
\bibliographystyle{aasjournalv7}



\end{document}